\documentclass[12pt,preprint]{aastex}

\newcommand{\degree}{\mbox{$^{\circ}$}}

\lefthead{Allers, Kessler-Silacci, Cieza \& Jaffe}
\righthead{Low-mass Brown Dwarfs with Mid-IR Excesses}

\begin{document}

\title{Young, Low-mass Brown Dwarfs with Mid-Infrared Excesses}

\author {K. N. Allers, J. E. Kessler-Silacci, L. A. Cieza, D. T. Jaffe}
\affil{Department of Astronomy,
University of Texas at Austin, Austin, TX 78712-0259}
 
\begin{abstract}
We have combined new I, J, H, and Ks imaging of portions of the 
Chamaeleon~II, Lupus~I, and Ophiuchus star-forming clouds with 
with 3.6 to 24 $\mu$m imaging from the 
Spitzer Legacy Program, ``From Molecular Cores to Planet Forming Disks'', to 
identify a sample of 19 young stars, brown dwarfs and sub-brown dwarfs 
showing mid-infrared excess emission.  The resulting sample includes sources 
with luminosities of 0.5~$>$~log(L$_{\ast}$/L$_{\odot}$)~$>$~-3.1. 
Six of the more luminous sources in our
sample have been previously identified by other surveys for young stars 
and brown dwarfs.  
Five of the sources in our sample have nominal masses at or below the 
deuterium burning limit (12~M$_{\rm J}$).
Over three decades in luminosity, our sources have an approximately constant  
ratio of excess to stellar luminosity.  We compare 
our observed SEDs to theoretical models of a central source with a passive 
irradiated circumstellar disk and test the effects of disk inclination, 
disk flaring, and the size of the inner disk hole on the strength/shape of 
the excess.  The observed SEDs of all but one of our sources are well fit 
by models of flared and/or flat disks.  
 
\end{abstract} 
 

\section{Introduction}
Young, free-floating objects with masses comparable to those of extrasolar 
planets, or sub-brown dwarfs 
(M~$<$~12 M$_{\rm Jupiter}$, hereafter M$_{\rm J}$), 
have been difficult to find and even more difficult to confirm.  
To date, searches for planetary-mass brown dwarfs have focused on dense 
young stellar clusters with large extinctions in order to 
eliminate contamination by background objects.  
Several surveys have reported sources with masses possibly below 12 M$_{\rm J}$ in 
Orion \citep{lucas05,zapatero02,lucas01}, but the intrinsic 
faintness of these objects, combined with the large distances to the sources, 
makes it difficult to confirm spectroscopically the low gravity and hence 
young age and low mass of the objects.  S Ori 70, originally thought to 
be a young, 3 M$_{\rm J}$ object in the $\sigma$~Orionis cluster \citep{zapatero02},
may actually be an older, more massive field brown dwarf 
\citep{burgasser04}.  Recently, \citet{chauvin04} directly detected a 
possible young ($\sim$8~Myr), low-mass ($\sim$5~M$_{\rm J}$) companion to a 
brown dwarf.  The faintness of this source (K=16.93), in combination with 
contamination from its parent star (0.8\arcsec\ away), 
makes it difficult to obtain the high resolution, high S/N 
spectra necessary to characterize the temperature and gravity of the object. 
The lack of confirmed free-floating, young, planetary mass 
objects means that 
the properties of these objects remain relatively unknown and makes finding 
them, based on optical and near-IR photometry alone, very difficult.  

An unambiguous way to confirm the youth of brown dwarfs is by detection 
of excess emission from a circumstellar disk. \citet{haisch01} found that 
half of the stars in 
young clusters lose their inner disks before they reach 
$\sim$3 Myr old.  
Given the similarity of 
disk fractions for young brown dwarfs and T~Tauri stars 
\citep{luhman05a,jayawardhana03,liu03}, brown dwarf disks may have 
inner disk lifetimes similar to those of disks around T~Tauri stars.
A substantial fraction of stellar and sub-stellar 
objects in star forming clusters 
show excess emission above their 
photospheres at infrared wavelengths, 
presumably due to a circumstellar disk \citep{liu03, wilking04, 
lada04, muench01}. 
Ground-based surveys searching for direct evidence of circumstellar disks 
have been limited to searching for L$^{\prime}$ and K band excesses.  These 
surveys have found circumstellar disks around objects 
with inferred masses as 
low as 15~M$_{\rm J}$ \citep{liu03, jayawardhana03}.  
The low luminosity of sources with masses below 15~M$_{\rm J}$ cannot passively 
heat enough material in the disk to create substantial excess emission in the 
near-IR.  Models of irradiated circumstellar disks around sources with 
masses below 15 M$_{\rm J}$ indicate that 
emission in the optical, near-IR and L bands is 
dominated by contributions from the object's photosphere, whereas at 
mid-IR wavelengths, emission from the disk begins to exceed the photosphere  
\citep{natta02, walker04}.  Recent 
studies of disks around low mass objects \citep{luhman05b, luhman05, natta02} 
have begun to look at the nature of disks around brown dwarfs with masses 
approaching and below the deuterium burning limit \citep[12~M$_{\rm J}$;][]{saumon96} and 
luminosities as low as log(L$_{\ast}$/L$_{\odot}$) $\sim$ -3.2.  
The sensitivity of the Spitzer Space Telescope 
\citep[hereafter \emph{Spitzer},][]{werner04} in the mid-IR, 
in combination with the sensitivity of current optical and near-IR imagers 
makes it possible to detect excesses around sources with such low 
luminosities, allowing us to search for sub-brown dwarfs with 
circumstellar material.

There are several unanswered questions concerning circum-brown dwarf disks.
What is the lowest mass object that can harbor a 
circumstellar accretion disk?  
If brown dwarfs and planetary mass objects are products 
of the same cloud-fragmentation and collapse process 
that forms higher mass objects, one would expect to find circumstellar disks 
around even the lowest mass cloud members.  
Given the low luminosity of the central source and the low gravitational 
force that the central object exerts on the disk, 
is it plausible for circumstellar 
material around sub-brown dwarfs to have similar structures and 
lifetimes as disks around T~Tauri stars?  \citet{luhman05} suggest that excess 
emission around OTS44, one of the 
lowest-mass objects known to have a circumstellar disk, 
cannot be explained by passive heating of the disk and 
requires additional heating from viscous dissipation, implying 
an accretion rate larger than observed for 25 M$_{\rm J}$ objects and more 
in line with accretion rates around T~Tauri stars \citep{muzerolle05}.  If 
low-mass brown dwarfs have circumstellar material, can they in turn form 
planets?  The recent detection of a possible low-mass companion to an 
M8.5 type brown dwarf \citep[$\sim$25 M$_{\rm J}$ at 8 Myr;][]{chauvin04} sharpens 
interest in this possibility.

The capabilities of \emph{Spitzer} 
provide a unique opportunity to search for disks around objects 
with masses extending into the planetary-mass regime.  
The Spitzer Legacy Program, ``From Molecular Cores to Planet Forming Disks'' 
\citep[hereafter c2d;][]{evans03}
provides IRAC and MIPS fluxes for several star forming regions.  
While mid-IR data from IRAC and MIPS 
are necessary for detecting excesses around brown dwarfs, 
optical and near-IR photometry are needed to probe the photosphere and derive 
object extinctions, luminosities, temperatures and masses.
To search for low-mass brown dwarfs with circumstellar disks, we have 
undertaken a large near-IR survey of selected areas of the c2d survey.  We 
derive the properties of the 
central source (luminosity and extinction) from our 
I-band and near-IR observations and look for excess emission 
in the IRAC and MIPS bands above that 
expected from the photosphere.  The resulting sample 
includes young stars, brown dwarfs and sub-brown dwarfs 
with a range of luminosities (0.5 $>$ 
log(L$_{\ast}$/L$_{\odot}$) $>$ -3.1), and nominal masses as low as 6~M$_{\rm J}$.
We then compare our measured excesses to model predictions of an irradiated 
circumstellar disk, and discuss the differences in circumstellar material as a 
function of central source luminosity.   

\section{Observations and Data Reduction}
\subsection{Survey Area Selection}
\begin{figure}
\plottwo{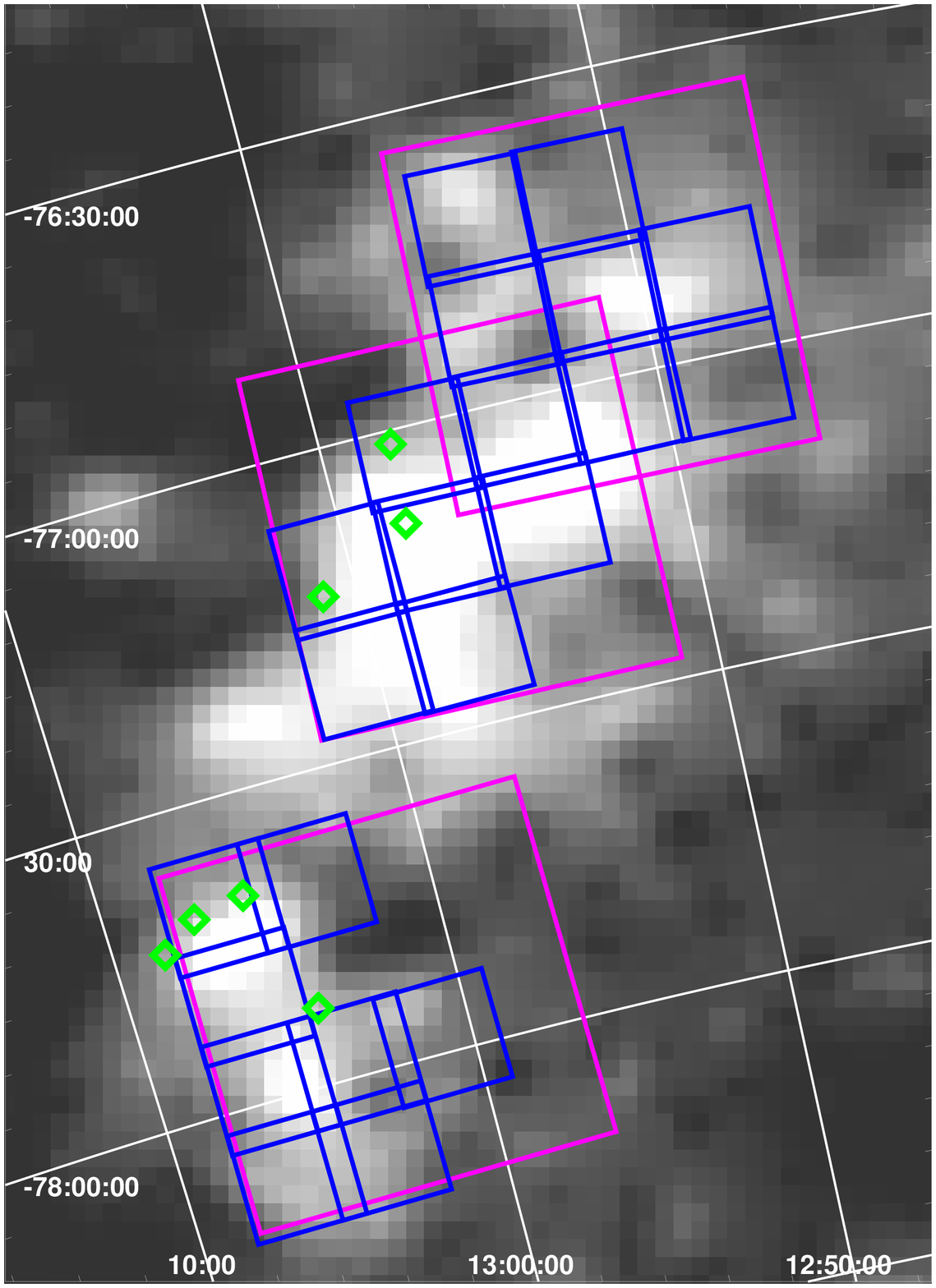}{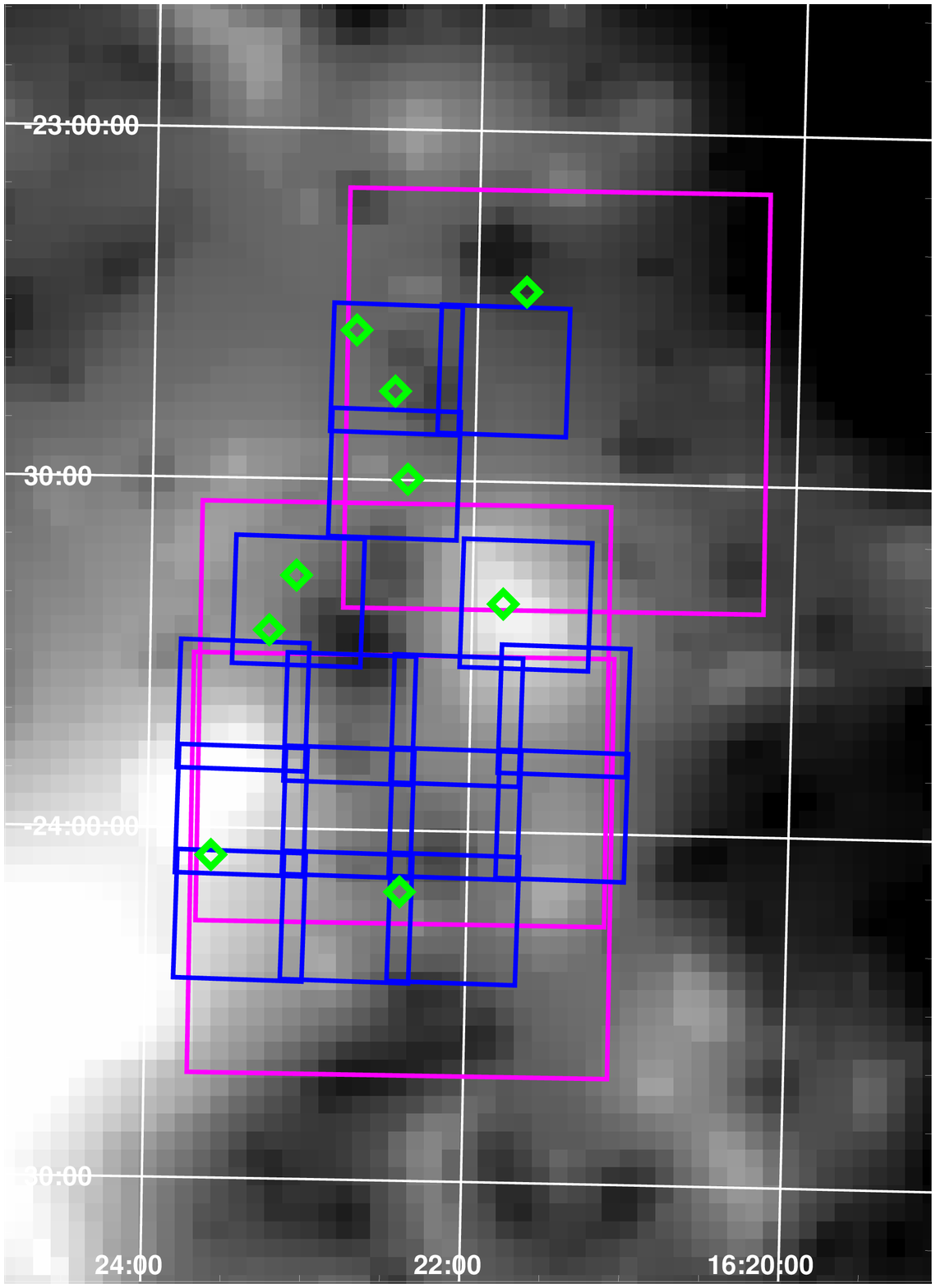}
\caption[1]
%>>>> use \label inside caption to get Fig. number with \ref{}
{\label{fields} The locations of our survey areas in Chamaeleon~II (left) and 
Ophiuchus (right) superposed on extinction maps from \citet{cambresy99}.  
The magenta boxes show the positions of our MOSAIC II 
(I band) fields, and the blue  boxes show the locations of our ISPI (near-IR) 
fields.  The green diamonds show the locations of sources \#1 to \#16 
in our sample of objects showing mid-IR excess emission.  While observing, 
we dithered ISPI by up to 45\arcsec, increasing the size of our ISPI fields 
in some cases, as evidenced by the 2 green diamonds that fall outside of the 
displayed ISPI field boundaries. 
The A$_V$'s 
shown in the maps range from 0.1 to 5.4 in our ISPI surveyed regions of 
Chamaeleon~II and 1.3 to 7.3 in our ISPI surveyed areas of Ophiuchus. 
}
\end{figure}

\begin{figure}
\epsscale{0.6}
\plotone{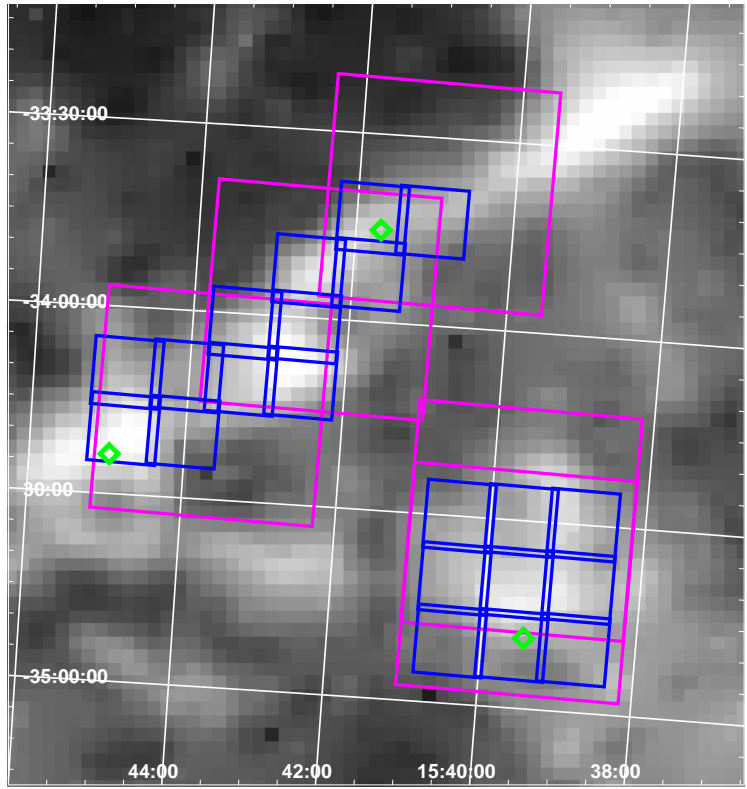}
\caption[1]
%>>>> use \label inside caption to get Fig. number with \ref{}
{\label{fieldslup} 
The locations of our survey areas in Lupus~I 
superposed on an extinction map from \citet{cambresy99}.  
The magenta boxes show the positions of our MOSAIC II 
(I band) fields, and the blue boxes show the locations of our ISPI (near-IR) 
fields.  The green diamonds show the locations of sources \#17 to \#19 
in our sample of objects showing mid-IR excess emission.
The A$_V$'s 
shown in the map range from 1.2 to 4.8 in our ISPI surveyed regions. 
}
\end{figure}

The c2d program has imaged 16 square degrees toward 5 
galactic star forming regions using both IRAC and MIPS.  
For our follow-up survey in I, J, H, and Ks, we chose areas of the c2d 
survey in the Ophiuchus, Chamaeleon~II, and Lupus~I clouds having
modest extinction as shown in Figure~\ref{fields} 
\citep[A$_V<7.5$ magnitudes;][]{cambresy99}.  To select regions with 
evidence for recent star formation, we deliberately chose areas containing or 
close to 
concentrations of T Tauri stars known from H$\alpha$ surveys 
\citep{wilking87,hartigan93,schwartz77}.
The positions of our fields in the
Ophiuchus, Chamaeleon~II, and Lupus~I molecular clouds are shown in 
Figures \ref{fields} and \ref{fieldslup}.
The total area covered and average limiting magnitudes 
of our survey are shown in Table \ref{limits}. 

\subsection{Near-Infrared Imaging}
We obtained J, H, and Ks images using the Infrared Sideport Imager 
\citep[ISPI;][]{vanderbliek04,probst03} on the CTIO Blanco 4m telescope on 
2003 May 16 \& 17 and 2004 May 1--5. 
The 2048 $\times$ 2048 Hawaii II InSb array in ISPI has a 
0.305\arcsec\ pixel$^{-1}$ plate scale.  
K-band FWHM seeing measurements ranged from 0.6\arcsec\ to 2.0\arcsec .
Typical on-source integration times 
were 10 minutes in each band (1 minute integrations taken in a 
10-point random dither pattern with maximum shifts of 45\arcsec ).  
On-source integration times were varied with seeing to maintain uniform 
sensitivity limits (Table \ref{limits}).
The total integration time in any one band varied from 5 to 30 minutes.
After flat fielding the raw frames, we masked out saturated pixels, 
subtracted the median from each frame, and created a sky frame by median 
combining 
the images in a 10 point dither.  After subtracting the sky from each frame, 
we used IRAF\footnote{IRAF is distributed by the National Optical 
Astronomy Observatories,  which are operated by the Association of 
Universities for Research in Astronomy, Inc., under cooperative 
agreement with the National Science Foundation.} 
and the IMCOADD task 
(part of the GEMTOOLS package) to derive the position shifts between frames 
and create an averaged, cosmic-ray-event cleaned image.  
For fields with more than 1000 matched sources with the USNO-A2 catalog 
\citep{monet98}, we created a map of average 
field distortion by fitting 4th order 
polynomials with full cross terms in the x and y pixel directions.  We 
combined all of the field distortion maps for each band, 
and applied the result to our near-IR images.
After correcting for field distortion, we 
established the world coordinate system 
(WCS) for our on-source images by cross-correlating our 
images with the USNO-A2 catalog.  Typical residuals for our WCS fit are 
0.2\arcsec.  We found sources in our fields using the Source Extractor 
package \citep{sextractor}.  We then performed aperture photometry 
using the PHOT task (part of the IRAF APPHOT package) 
with an aperture radius of 1.1 times the median FWHM of objects found in the 
image.  The 1 $\sigma$ uncertainty for our aperture photometry was 
calculated by the PHOT task. ISPI uses J, H, and Ks filters from the Mauna Kea 
Observatory (MKO) filter set \citep{simons02}.  
To set the zero point for our photometry, we matched sources found 
in our survey to sources that are at least 20$\sigma$ detections in 
the 2MASS catalog.  We transformed the matched 2MASS source magnitudes to the 
MKO system using the transformation equations found in the 2MASS Explanatory 
Supplement.  For an object with a J-Ks color of 1.0 magnitude and a J-H color 
of 0.5 magnitudes, the uncertainties associated with the filter 
transformations were 0.011, 0.010 and 0.007 magnitudes in J, H, and Ks.
The RMS scatter about the zero point of our sources from 
the zero point solution 
determined from the 2MASS catalog varied from 0.007 magnitudes in the 
best matched fields to 0.02 magnitudes in the worst matched fields.  
The 2MASS explanatory supplement claims calibration uncertainties of 0.011, 
0.007, and 0.007 magnitudes in J, H, and Ks respectively.  Therefore, we assign 
a conservative 0.03 magnitude calibration error to our near-infrared colors.
To ensure that our catalog includes accurate near-IR photometry of objects 
saturated in our ISPI images, we combined our near-infrared catalog and any 2MASS sources in our fields, 
after transforming the 2MASS magnitudes to the MKO filter system.

\subsection{Optical Imaging}
We obtained our I band images using the MOSAIC II imager \citep{muller98} 
on the Blanco 4m
telescope at CTIO on 2004 March 26 \& 27.  
The I band filter used on MOSAIC II is 
centered at 0.805 $\mu$m and has a FWHM of 0.150 $\mu$m.
The positions of the 
36\arcmin\ $\times$ 36\arcmin\ fields are shown in Figure 
\ref{fields}.  The plate scale of the 
eight 2048 $\times$ 4096 SITe CCDs is 0.27\arcsec\ pixel$^{-1}$.  
I-band FWHM seeing measurements were 
$\sim$0.7\arcsec\ on the night of 2004 March 26 and $\sim$1.3\arcsec\ on 
2004 March 27.
During our observing run, 
one of MOSAIC II's CCDs was inoperable.   By carefully selecting our dither 
pattern and field placement, we made sure that all of our fields observed by 
ISPI were evenly covered.  On the night of 2004 March 26, we took 300~s 
exposures over a 5 point dither pattern for 1500~s of on source 
integration time.  Since the seeing was worse on the night of 2004 March 27, 
we took 300~s exposures over a 10 point dither pattern for 3000~s of on source 
integration time, in order to obtain roughly uniform sensitivity for data 
taken either night.   We also took individual 30~s exposures of each field in 
order to achieve higher dynamic range in our catalog.  
The 10$\sigma$ limits for our I-band data are shown in Table \ref{limits}.
We used the MSCRED package in IRAF to flat-field the images, 
create bad pixel masks, derive the WCS for each frame, 
create a single images from the mosaic, and combine the images into 
a final averaged image, as outlined 
by \citet{mscred}.  We used Source Extractor \citep{sextractor} 
to find sources in our fields.   MOSAIC II's large field of view 
means that the FWHM of the PSF can vary by as much as 20\% across the field.
Because of this variation, we used larger 
aperture radii than the 1.1 times the FWHM used for our near-IR data 
reduction.  We obtained aperture photometry using 
APPHOT for an aperture radius of 1.25\arcsec\ with inner 
and outer sky radii of 5.0\arcsec\ and 10.0\arcsec\ 
respectively for data taken on 
2004 March 26, and an aperture radius of 1.50\arcsec\ 
with inner and outer sky radii of 6.0\arcsec\ and 12.0\arcsec\ 
respectively for data taken on 2004 March 27.  Photometric zero 
points were obtained by observing Landolt standard fields \citep{landolt92} 
and adding a linear term to correct for airmass as well as an offset for the 
aperture correction.  The RMS scatter about the zero point correction was 
0.007 magnitudes for standard fields observed on 2004 March 26 
and 0.006 magnitudes for fields observed on 2004 March 27. 
Comparing data taken in overlapping fields, we found 
that our positions agreed to better than 0.25\arcsec\ while our 
magnitudes agreed to better than 0.05 magnitudes. 
We therefore assign a positional uncertainty of 0.25\arcsec\ 
and a photometric error of 0.05 magnitudes to our I band photometry.

\subsection{C2D Observations}
                                                                               
As part of the c2d Legacy Project \citep{evans03}, 
\emph{Spitzer} has mapped 8.0 square 
degrees of the Ophiuchus molecular cloud, 1.1 square degrees of the 
Chamaeleon~II molecular cloud and 2.4 square degrees of the Lupus molecular 
clouds with IRAC \citep[3.6, 4.5, 5.8, and 8.0 $\mu$m;][]{fazio04} and MIPS 
\citep[24 $\mu$m;][]{rieke04}. The IRAC maps
consist of two epochs, separated by several hours, each 
with two dithers of 12~s observations (48~s total). The second 
epoch observations
were taken in the High Dynamic Range mode which include 0.6~s
observations before the 12~s exposures, allowing photometry
of both bright and faint stars. MIPS observations were
taken in the fast scan mode, also in  two different epochs of 15~s
exposures each. The average 10 $\sigma$ sensitivities of the c2d
survey of molecular clouds are shown in Table \ref{limits}.  Detailed 
descriptions of the MIPS and IRAC observing procedures used by c2d 
can be found in \citet{young05} and \citet{harvey06}. 

\subsection{IRAC and MIPS Photometry}

We searched preliminary c2d IRAC and MIPS point source catalogs for fluxes
of brown dwarf candidates selected in the near-IR (see \S 3.1), and used these 
fluxes when available. The c2d
catalogs were produced using the c2d mosaicking/source extraction
software, c2dphot \citep{harvey06}, 
which is based on the mosaicking
program APEX developed by the \emph{Spitzer} Science Center and the source
extractor Dophot \citep{schecter93}. Since the
preliminary c2d catalogs in the archive 
are complete down to only $\sim$ 20 $\sigma$
detections (see \citet{evans05} for a detailed description of the c2d
data products), we had to extract our own fluxes from the c2d images for many
of our sources (especially at 5.8 and 8.0 $\mu$m). To ensure consistency, we
extracted fluxes not present in the c2d catalogs using the c2dphot
software in a mode that allows us to obtain fluxes (and upper limits) of
low S/N sources with known positions. In this mode, c2dphot calculates the
fluxes by fitting a PSF to fixed positions. The only free
parameters are the background level and peak intensity. The flux
uncertainty is calculated from the goodness of the fit. The input
coordinates were taken from our near-IR observations which, in general,
show a coordinate agreement of $\lesssim$~0.3\arcsec\ with c2d
sources.  We consider $\ge$~5~$\sigma$ detections to be real. For all detections, however, we examine the images using the IRAF IMEXAMINE task, 
and make sure the object is a point-source and well detected above the 
background.
We note that the error
in the photometry of high S/N sources is dominated by the absolute
calibration uncertainty for the c2d IRAC and MIPS catalogs \citep[$\sim$15$\%$;][]{evans05}.

\section{Results}
\subsection{Source Selection}

In our Chamaeleon~II, Ophiuchus, and Lupus~I 
fields, we detected $\sim$120,000 sources at 
5 $\sigma$ or better in the I-band and all three near-IR bands.
From our full catalog, we wish to select only objects that 1) 
are likely young stellar objects and 2) show evidence of mid-IR
excess emission from 
circumstellar material.  To create our final list of young sources with mid-IR 
excess emission, we must weed out extragalactic objects, background stars, 
and foreground stars.

Even in the presence of foreground reddening, extragalactic objects 
have near-IR colors that allow them to be distinguished 
from brown dwarfs.  
The Munich Near-Infrared Cluster Survey \citep{drory01} covers $\sim$1 square 
degree at high galactic latitudes to limiting I, J, and K magnitudes of 
22.4, 21.0 and 19.5 respectively.
Galaxies from \citet{drory01} typically have I-J colors of $\sim$1.0.
For young M-type objects, 
later spectral types, and hence lower effective temperatures 
correspond to redder intrinsic 
I-J colors \citep{briceno02, luhman04c}.  The trend of redder I-J colors with 
later spectral type is also observed for M, L and T spectral type 
field brown dwarfs 
\citep{dahn02}.  The average I-J color of young 
M6 objects \citep{briceno02, luhman04c} is 2.42.
The number counts of galaxies increase as one moves to fainter magnitude bins
\citep{drory01}. Thus, we require strict selection criteria (redder I-J colors)
for faint objects, and loosen our selection criteria for brighter objects.  

\begin{figure}
\plotone{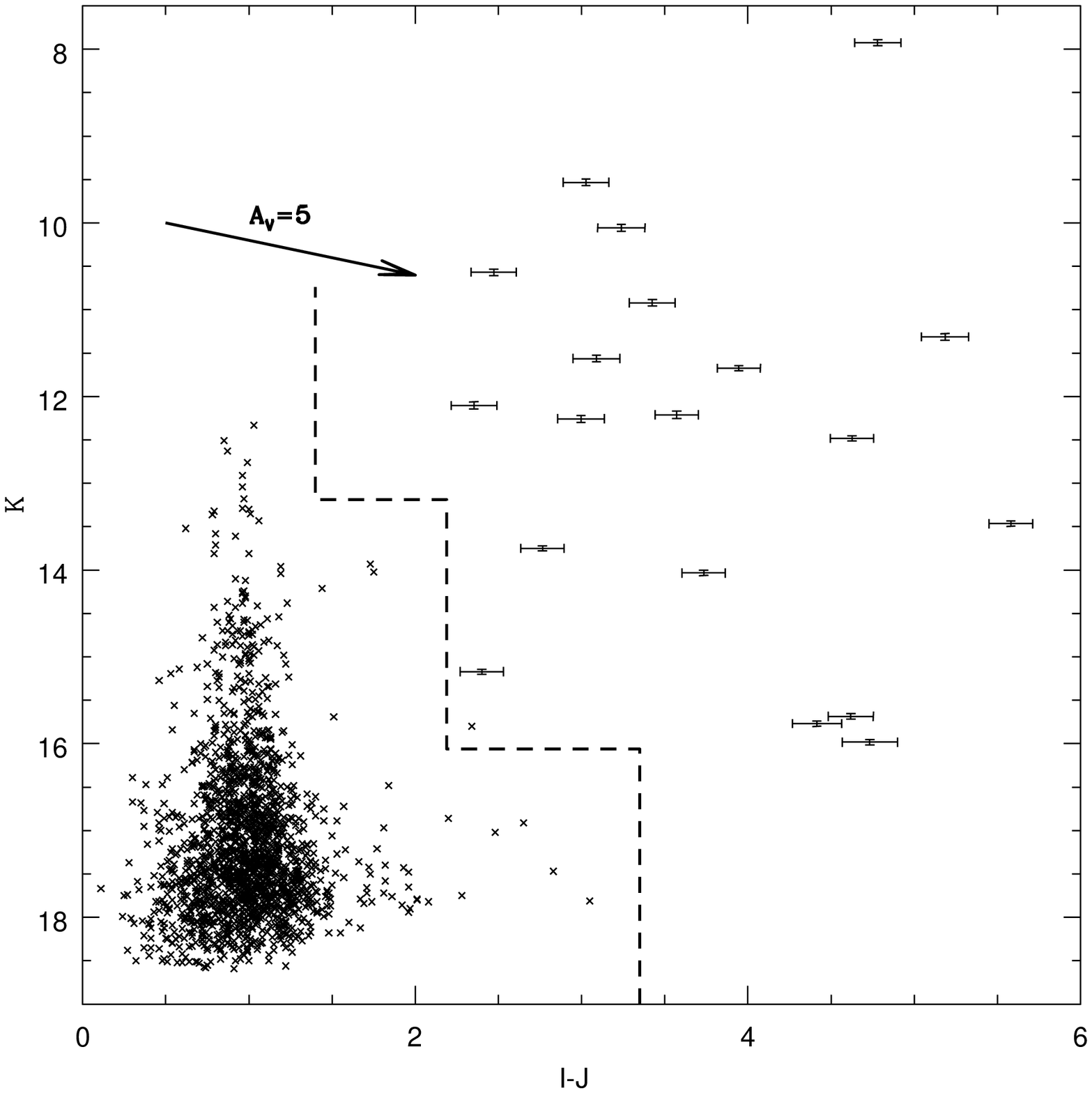}
\caption[1]
{\label{ij_k} Our color-magnitude source selection criteria.
The points with error bars are objects from Table \ref{sources}.  
The X's show galaxies from \citet{drory01}.
Objects to the right of the dashed line: 
1) with K magnitudes greater than 9.81 + the distance modulus the cloud 
($\mu_d =$~6.25 for Chamaeleon~II, 5.48 for Ophiuchus, and 5.88 for Lupus~I) 
and I-J colors redder than 3.35, 2) objects 
with K~$<$~9.81~+~$\mu_d$ and I-J~$<$~2.19, 3) objects with 
K~$<$~6.94~+~$\mu_d$ and I-J~$>$~1.40, and 4) objects with 
K~$<$~4.49 meet our selection criteria.}
\end{figure}

\begin{figure}
\plotone{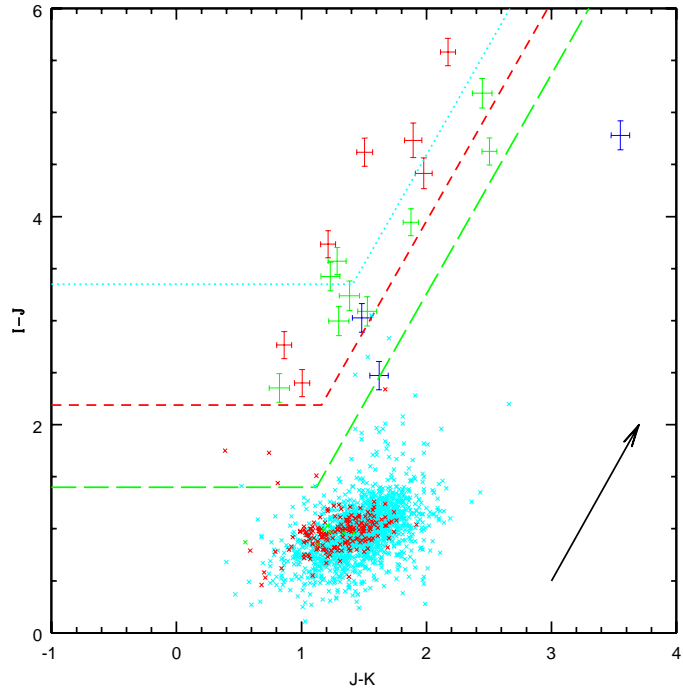}
\caption[1]
{\label{jkij} Our color-color source selection criteria.
The points with error bars are objects \#1 to \#16 in Table \ref{sources}.  
The X's show galaxies from \citet{drory01}.
The cyan sources have K magnitudes less than 9.81 + the distance modulus to 
Chamaeleon~II ($\mu_d=$ 6.25).  
These sources must lie above the cyan dotted line to meet our 
selection criteria.  Similarly, to meet our selection criteria, 
the red sources (K magnitudes between 9.81 and 6.94 + $\mu_d$) must lie 
above the red dashed line and the cyan sources 
(K magnitudes between 6.94 and 4.49 + $\mu_d$) must 
lie above the cyan long-dashed line.  
The blue sources have K magnitudes brighter than 4.49 + $\mu_d$ 
and have no color criteria for selection.}
\end{figure}

We initially cut our sample based on the I, J, H, Ks colors and magnitudes 
of the objects.  
We start with 120,000 sources detected at $>5~\sigma$ in I, J, H, and Ks, and 
apply different selection criteria depending on whether the object's K 
magnitude is fainter than expected for a young M9 object, 
between the expected K magnitudes for young M6 and young M9 objects, between 
the expected K magnitudes for young M3 and young M6 objects, or brighter 
than the K magnitudes of young M3 objects.  We determine the colors and 
magnitudes of young M objects from samples in Taurus and 
Chamaeleon~I \citep{briceno02,luhman04c}.  Our near-IR 
source selection criteria 
are illustrated in Figures~\ref{ij_k} and \ref{jkij}.
The average absolute K magnitude of young M8.5 to M9.5 objects in 
Taurus and Chamaeleon I is 8.01 \citep{briceno02,luhman04a}.  
The bluest I-J color of a young M8.5 to M9.5 
object with an absolute K magnitude less than 8.01 is 3.35.  
Thus, for objects with observed K magnitudes fainter than 8.01 plus 
the distance modulus to the cloud, $\mu_d$ (Table \ref{clouds})
we select only objects with I-J$>$3.35.  
To ensure that we do not 
add reddened galaxies to our sample, we deredden the objects to the average 
J-K color (1.41) of M8.5 to M9.5 young brown dwarfs, and select objects with 
a dereddened I-J$>$3.35 (including photometric errors).
We use A$_{\lambda}$/A$_V$ values for our I, J, and Ks filters (0.56, 0.26, 
and 0.12 respectively) from the Asiago Database on Photometric 
Systems\footnote{http://ulisse.pd.astro.it/Astro/ADPS/} \citep{adps}.
For objects 
which range in magnitude between the average absolute K magnitude for 
young M8.5 to M9.5 dwarfs
reddened by A$_V$=15 and the average absolute K magnitude of young 
M6 objects (9.81 and 5.14 respectively) + $\mu_d$, 
we require I-J$>$2.19 when 
dereddened to a J-K of 1.16 (appropriate for young M6 objects). 
For objects with K magnitudes between the average of young M6 objects
reddened by A$_V$=15 and the average absolute K magnitude of young 
M3 stars (6.94 and 2.69 respectively) + $\mu_d$, 
we require I-J$>$1.40 when 
dereddened to a J-K of 1.12 (appropriate for young M3 objects).
Objects brighter than the 
intrinsic K magnitude of young M3 objects + $\mu_d$ 
are few in number (63), and none of the galaxies from \citet{drory01}
are this bright, so we include all of these sources in our initial sample.  
For each object meeting our color and magnitude
selection criteria, we examine the object's PSF to ensure that is a 
point source.
5853 objects toward Ophiuchus, 1504 objects toward 
Chamaeleon~II, and 4977 objects toward Lupus~I 
meet our initial selection criteria.
None of the $\sim$3300 galaxies observed by \citet{drory01} that are within 
the detection limits of our near-IR survey (Table \ref{limits}) 
meet our near-IR selection criteria for young brown dwarfs, 
even if we redden the galaxies by A$_V$=5.  We also compared templates of 
all the galaxy types fit to objects in the Lockman hole 
(Polletta et al., in preparation) by red shifting them 
from z=1 to 4 and reddening them from A$_V$=0 to 10.  
None of the Polletta et al. model 
templates which might appear as point sources 
in our I-band images meet our selection criteria. 

For the sources meeting our near-IR selection criteria, we make an additional 
cut by looking for mid-IR excesses.
At the galactic latitudes of 
the clouds in our survey (b~$\sim$~18, 16 and -14 for Ophiuchus, Lupus~I, 
and Chamaeleon~II
respectively), there are no known star-forming regions behind our clouds.  
By looking for objects with evidence 
of emission in the mid-IR significantly in excess of the photospheric emission 
extrapolated from the near-IR colors (i.e., 
evidence for a disk or circumstellar material), 
we will ensure the young age (and hence 
cloud membership) of candidates in our sample, effectively eliminating 
any field-population 
background or foreground stars or brown dwarfs, which will not have excess 
emission.
We obtain mid-IR fluxes for the 12334 objects meeting our near-IR selection 
criteria using the method described in \S2.5.
Using the near-IR colors of objects in our sample, we fit 
reddening simultaneously with the observed I, J, H and Ks 
colors of young M-type objects in Chamaeleon I and Taurus 
\citep{briceno02,luhman04c} as described in \S3.2.  To estimate the expected 
mid-IR emission from the photosphere of our best fit young object, we use the 
observed IRAC colors of field brown dwarfs \citep{patten04}.
If an excess exists at [5.8] and [8.0] of greater than 3 $\sigma$ above the 
best fit young M-type object colors, we examine the near and mid-IR images 
to make sure that the object is a point source, since IRAC bands 3 and 4 
suffer from substantial nebulosity.  Table \ref{sources} 
contains the observed I, J, H, Ks, and IRAC magnitudes for the 19 sources in 
our survey that have colors and magnitudes meeting our selection criteria 
and which appear as point sources in the I, J, H, Ks and IRAC images.  
Sources \#1--\#19 are detected in all of the 
IRAC bands as well as in MIPS1, with excess emission above that expected from 
the photosphere in IRAC3, IRAC4, and MIPS1.  
Flux densities from 3.6 to 24 $\mu$m for 
objects \#1 to \#19 are shown in Table \ref{iracsources}.  
The SEDs of 18 of the sources comprising our 
sample are shown in Figures \ref{fig:sed1} and \ref{fig:sed2}, 
where the open circles 
show the observed data and the filled circles show the extinction-corrected 
data; we omitted source \#3, the brightest object in our sample.

\begin{figure}
\hspace{1cm}
\includegraphics[angle=90,width=6in]{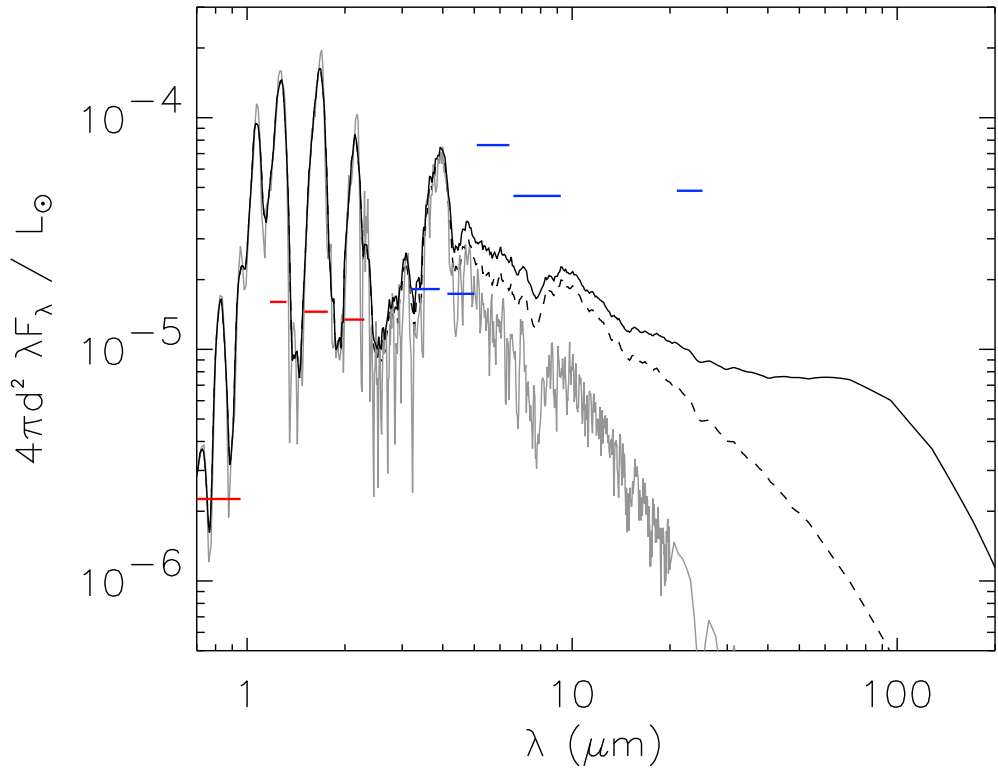}
\figcaption{Theoretical SED of a 1~Myr old, 2 M$_{\rm J}$ sub-brown dwarf at the 
distance to Ophiuchus \citep[125 pc;][]{degeus89}).  The grey line shows the 
theoretical photospheric emission of a 1~Myr old, 
2 M$_{\rm J}$ object \citep[1300 K, log(g)=3.5;][]{baraffe03, allard01}.  
Theoretical SEDs for a 2 M$_{\rm J}$ object with a flared (solid line) 
and a flat (dashed line) circumstellar disk are also shown (See \S 5 
for details of the disk models).  We chose parameters for our disk models 
($i=$~0\degree, and $R_i=R_{\ast}$) that produce 
the largest mid-IR fluxes.  The horizontal lines denote the band widths 
and 10~$\sigma$ sensitivity limits of our I, J, H, and Ks survey toward 
Ophiuchus along with 10~$\sigma$ IRAC and MIPS~1 sensitivity limits of the 
c2d survey toward Ophiuchus.  Though a 1~Myr old, 2 M$_{\rm J}$ sub-brown dwarf 
would be easily detected by our near-IR survey, 
it is below the c2d sensitivity limits in IRAC~3, IRAC~4, 
and MIPS~1, even with excess emission 
from a flared disk. \label{2mj}}
\end{figure}

None of the 19 sources in our sample have K magnitudes that fall in our 
faintest magnitude bin (8.01 plus A$_V$=15 + $\mu_d$) for selection criteria. 
The faintest observed K magnitude of our sample is 
3 magnitudes brighter than our 10 $\sigma$ limit.  The c2d survey does not 
have the sensitivity in IRAC bands 3 and 4 to detect objects with fainter 
intrinsic K magnitudes.  Figure \ref{2mj} shows the theoretical SED for a 
1~Myr old, 2~M$_{\rm J}$ 
sub-brown dwarf \citep{baraffe03,allard01} along with models of two 
possible circumstellar disks 
(\S~5 details our disk modeling procedure).  
Even though this object would 
be detected easily in our near-infrared survey, it is fainter from 
5.8 to 24~$\mu$m than the 10~$\sigma$ c2d IRAC and MIPS limits.  
1633 sources in our survey have 0.8 to 4.5 $\mu$m colors which meet the 
selection criteria for objects fainter than M9, but are not detected in 
IRAC3, IRAC4 and MIPS1.  Some of these objects may be very low-mass 
objects with disks.

We have tested our source selection in several ways.  Examination of the 
colors of young brown dwarfs with disks shows that we can recover objects 
of this type.  
\citet{natta02} successfully fit excess emission (as detected with 
ISOCAM) around young 
brown dwarfs in the core of Ophiuchus 
with models of emission from a circumstellar disk.  
More recently, \citet{luhman05b, luhman05} detected (with IRAC) 
and modeled circumstellar disks around two brown dwarfs in Chamaeleon I (Cha~1109-7734 and OTS44) with 
masses of $\sim$8 and $\sim$15 M$_{\rm J}$.  
Using I-band fluxes from the 
the literature, 2MASS or published near-IR magnitudes 
(transformed to the MKO system), and 
IRAC fluxes from c2d (Allen et al. 2006, in preparation) or 
\citet{luhman05b,luhman05}, 
we have applied our near-IR selection criteria to these 8 objects and then 
searched for mid-IR excesses as outlined above.
Cha~1109-7734, OTS44, and all of the \citet{natta02} objects meet our selection criteria.
Six of the more luminous sources among the 19 
in our sample have been identified 
in other surveys searching for young stars and brown dwarfs, which also lends 
credibility to our selection criteria.  Sources \#2 and \#3 were identified 
as young objects using DENIS J and Ks and ISOCAM 6.7$\mu$m and 14.3$\mu$m 
photometry \citep{persi03}.  Source \#3 was also detected in soft 
(0.1--0.5~keV) and hard (0.5--2.5 keV) X-ray emission with ROSAT 
\citep{alcala00}.  Sources \#4, \#6, \& \#7 were identified as low-mass 
T~Tauri or young brown dwarf candidates by \citet{vuong01}.  Based on their 
strong H$\alpha$ emission, sources \#4 
and \#10 have previously been identified as the T~Tauri stars 
Sz52 and WSB 14 respectively \citep{schwartz77,wilking87}.  Finally, 
follow-up near-IR spectra of 5 of our sources in Table \ref{sources} 
(\#1, \#2, \#5, \#11, \& \#14) have 
confirmed their identification as young brown dwarfs (Allers et al. 
in preparation).

\subsection{Estimating Extinctions and Luminosities}

To determine the extinction toward our sources, we 
deredden and fit them to the observed colors of young M-type objects in 
the Taurus and Chamaeleon I star forming regions \citep{briceno02,luhman04c}.  
We deredden our colors based on A$_{\lambda}$/A$_V$ values of 
the MKO filter and Bessell filter systems in the Asiago Database on Photometric
Systems \citep{adps}, for the 
R$_V$=3.1 extinction law of \citet{fitzpatrick99}.  For IRAC and MIPS 
wavelengths, we 
use the values of A$_{\lambda}$/A$_V$ for the R$_V$=3.1 extinction law of 
\citet{fitzpatrick99} averaged over the filter bandpasses.  
The values of 
A$_{\lambda}$/A$_V$ used are 0.56, 0.26, 0.17, 0.12, 0.06, 0.04, 0.03, 0.02, 
and 0.0 for I, J, H, Ks, [3.6], [4.5], [5.8], [8.0], and [24] respectively.
The R$_V$=5.0 extinction law provides 
the same A$_{\lambda}$/A$_V$ for near and mid-IR bands, 
and only slightly changes A$_{\lambda}$/A$_V$ for the I band (to 0.60).
Since near-IR colors get redder as one moves to later spectral types, several 
of our objects can be fit to either a late spectral type object with low 
extinction or an earlier type object attenuated by more dust.  To break this 
degeneracy, we can limit the range of spectral types that we 
use to fit our objects based on the 
K magnitudes of our sources.  Objects with absolute K magnitudes brighter than 
expected for a young M6 objects (5.14; \S 3.1) are fit to the colors of 
young objects with spectral types from M3 to M8, while objects with 
absolute K magnitudes fainter than expected for young M9 objects (8.01) 
are fit to the colors of young M6 to M9.5 objects.
Objects between these extremes are fit to the colors of young M3 to M9.5 
objects. Even these fairly loose restrictions are sufficient to 
leave only one dereddening solution for each object.  
Table \ref{luminosities} shows the calculated A$_V$ for each of our sources.
When we use this technique to derive A$_V$ for the Chamaeleon I objects 
Cha 1109-7734 and OTS44
\citep{luhman05b, luhman05}, our calculated extinctions 
(A$_V$=2 magnitudes for both sources) agree to 
within the uncertainties with 
the published extinctions (A$_V$ = 1 $\pm$ 1 mags).
We also calculated A$_V$'s for six brown dwarfs with disks 
from \citet{natta02}.  Our calculated A$_V$'s are on average 0.3 magnitudes 
higher than the published A$_V$'s in \citet{natta02}, which have 
an uncertainty of $\pm$1 mag; the largest variations were A$_V$=2 mags.
We assign an uncertainty in A$_V$ of $\pm2$ mags for our 
calculated extinction values.

We calculate the stellar luminosity by integrating the dereddened 
flux density from I through [3.6] over the frequency width of the filters.  
For wavelengths between bandpasses, we linearly interpolated the 
flux densities and integrated them over the frequency gap between the bands.
We calculate the bolometric luminosity based on the sum of the flux in and 
between the bands, assuming the distances in Table \ref{clouds}.
As a check, we applied our 
method for calculating bolometric luminosities to field brown dwarfs from 
\citet{golimowski04} using [3.6] fluxes from \citet{patten04}.  We find that 
our calculated luminosities typically agree with the \citet{golimowski04} 
values to better than 5\%.  Excess emission at [3.6] does not 
greatly affect our luminosity calculations.  
\citet{liu03} detected L$^{\prime}$ excesses around late-M-type 
young brown dwarfs.  
Including the largest K-L$^{\prime}$ 
excess (0.45 magnitudes) from \citet{liu03} in the [3.6] magnitude of the 
field dwarfs from \citet{patten04} increases the calculated luminosity by 
at most 7\% relative to the same source with no excess.  
The uncertainties in cloud distance (Table \ref{clouds}) correspond to 
uncertainties of 0.09, 0.12, and 0.18 dex in Chamaeleon~II, Lupus~I, and 
Ophiuchus respectively. 
The luminosity uncertainties resulting from uncertainties in distance, 
A$_V$ (0.15 dex), and our method of calculating 
the luminosity (0.02 dex) combine for a total uncertainty of $\pm$0.24 in
log L$_{\ast}$ for our Ophiuchus sources, $\pm$0.19 in log L$_{\ast}$ for 
our Lupus~I sources, and  $\pm$0.18 in log L$_{\ast}$ 
for sources in Chamaeleon~II.  Our sources have 
luminosities ranging from 0.5 $>$ log(L$_{\ast}$/L$_{\odot}$) $>$ -3.1 
(Table \ref{luminosities}).  The mean log(L$_{\ast}$/L$_{\odot}$) of our 
sample is $-1.8$.  
Four of the objects in our sample (\#1, \#5, \#12, \& \#17) 
have luminosities that are equivalent (to within the uncertainties) to 
the lowest luminosity young brown dwarf 
with mid-IR excess reported to date \citep{luhman05b}.

\subsection{Ages}
The ages of objects in our sample are relevant because we assume an age and 
use a theoretical isochrone for that age to estimate the mass and T$_{eff}$ 
of our objects using their luminosities.  
The ages of stars in the clouds in our survey have been 
determined mainly from the ages of T~Tauri stars and brown dwarfs with masses 
greater than 20 M$_{\rm J}$.
Placing young objects on the H-R diagram in order to determine their ages 
is difficult.  The distances to the 
sources are usually assumed to be similar to the distance to the parent cloud 
(which in itself is usually only accurate to $\pm$ 10 pc).   
The stellar luminosity is 
usually found by estimating the reddening and using a bolometric correction 
to a single dereddened band (usually I or J bands).   In addition, T$_{eff}$ 
is usually determined from a SpT-T$_{eff}$ relationship, and not determined 
empirically.  The age determinations in the literature depend strongly 
on the evolutionary models overlaid on the H-R diagram.  
For example, the models of \citet{dantona94} yield 
ages that are a factor of $\sim$3 lower than \citet{baraffe03} 
age estimates \citep[e.g.][]{cieza05}.
The combination of these uncertainties makes estimates of ages 
for individual sources very uncertain.  The definition of age itself differs 
from model to model.  Some models arbitrarily take the zero age to be the point
when a contracting, fully convective object starts to move down the Hayashi 
track \citep[e.g.][]{dantona94, baraffe98}, whereas other models define the birthline based on the conditions of the star at the onset of deuterium burning
\citep{palla99}.  Whether or not the same 
age can be applied to stars, brown dwarfs, and objects with masses below the 
deuterium burning limit remains unclear.  
In addition, the models do not include 
the effects of accretion through a disk.  Since we do not have spectral types 
for our objects, we must estimate their effective temperatures and masses 
from model isochrones \citep{baraffe03}
by assigning an age based on values found in the literature.  
Fortunately, uncertainties in age of 
$\pm$2 Myr (for objects with estimated ages of 
3 Myr) correspond to uncertainties 
in effective temperature of less than 100 K according to the isochrones of 
\citet{baraffe03}.  Mass uncertainties resulting from mis-estimation of ages 
are slightly larger.  The lowest luminosity sources in our survey 
(log($L_{\ast}/L_{\odot})\sim-3.0$) correspond to masses of $\sim$6~M$_{\rm J}$ for the 
1~Myr isochrone, and correspond to masses of $\sim$15~M$_{\rm J}$ for the 5~Myr 
isochrone.  

\subsubsection{Ophiuchus}
\citet{luhman99} find a median age of 0.3 Myr for the objects in the core of 
Ophiuchus using luminosities estimated from J band fluxes and the isochrones 
of \citet{dantona94}, in agreement with previous work \citep{greene95}.
\citet{prato03} find ages ranging from 0.4 to 1.5 Myr 
for several young binaries in the Ophiuchus region by estimating stellar 
luminosities as the luminosity of blackbody at the T$_{eff}$ of the star 
scaled to fit the observed J and H band fluxes and deriving an age from the 
isochrones of \citet{palla99}.  More recently, 
\citet{wilking05} find a median age of 2.1 Myr for sources 
surrounding the Ophiuchus cloud core.  Given the young age that most authors 
find for Ophiuchus,  
we adopt an age of 1 Myr for our sources in Ophiuchus, which is the youngest 
isochrone of the \citet{baraffe03} models.

\subsubsection{Chamaeleon~II}
\citet{alcala97} find a mean age of 1.3 Myr for T~Tauri stars in Chamaeleon~II,
based on stellar luminosities derived from measured I band fluxes and the 
evolutionary tracks of \citet{dantona94}, in agreement with 
earlier work \citet{hughes92}.  
Recently, \citet{cieza05} find an average age of 3.6 Myr for classical T~Tauri 
stars in Chamaeleon by placing objects on the HR diagram along with 
the evolutionary tracks of \citet{baraffe98} based on luminosities 
derived from the I band flux, and temperatures based on the spectral type of 
the objects.  Since we are using the \citet{baraffe03} isochrones in subsequent 
sections of this paper, and published isochrones are available at young ages 
of 1, 3, 5, and 10 Myr,
we adopt an age of 3 Myr for our sources in Chamaeleon~II.

\subsubsection{Lupus~I}
\citet{wichmann97} find a mean age 1.2 Myr for classical T~Tauri stars
in the Lupus star forming region, and a slightly older mean age of 3.1 Myr
for weak line T~Tauri stars by calculating stellar luminosities from blackbody
fits to the R band (with T$_{eff}$ given from the spectral type 
of the object), and getting ages of the stars from the evolutionary tracks of
\citet{dantona94}.  This is in agreement with \citet{hughes94}, who
find that Lupus~I and II contain 
a younger stellar population than Lupus~III or IV, 
and estimate an age of $\sim$1 Myr for classical T Tauri stars in 
Lupus~I and II.  For Lupus~I, we adopt an age of 1~Myr.

\section{Analysis}

\subsection{Masses}
Using ages of 1 Myr for Ophiuchus and Lupus~I and 3 Myr for Chamaeleon~II,
we estimate the effective temperatures and masses of our 19 objects by
matching the source luminosities to the isochrones of \citet{baraffe03}
and \citet{baraffe98}.
Our sample includes five 
sources with nominal masses below the deuterium burning
limit, the lowest luminosity
source in our sample (log(L$_{\ast}$/L$_{\odot}$)$=-3.1$) having
a mass possibly as low as 6~M$_{\rm J}$.  Sources in the Trapezium cluster having 
similar dereddened absolute H-band magnitudes to the faintest sources in our 
sample also have similar mass estimates \citep[8--11~M$_{\rm J}$;][]{lucas01}.  
Our mass estimates rely heavily on
the assumed ages for our sources.  If our sources are actually 10~Myr old, the
inferred mass of a log(L$_{\ast}$/L$_{\odot}$)~=~-3.1 source
would be 15~M$_{\rm J}$.  Additional uncertainty in the nominal masses of 
our sources originates in the evolutionary models themselves.  
\citet{hillenbrand04} find that 
the evolutionary models underpredict the dynamically determined 
masses of pre-main sequence stars by as much as 30\%.  For young brown dwarfs, 
no dynamical mass constraints are available in the current literature.  Thus 
the evolutionary models are relatively untested in the young, 
low-mass regime, and masses 
determined from the models are highly uncertain \citep{baraffe02}.

\subsection{Accretion vs. Reprocessing Luminosities}
If the mid-IR excess emission we observe can be attributed to a circumstellar 
disk, what is the heating mechanism for the disk?
For a flat, optically thick disk extending outward from the stellar surface, 
the luminosity due to reprocessing of light from the irradiated disk 
is\footnote{see \citet{hartmann98} and references therein for derivation of 
equations 1 and 2}
\begin{equation}
L_{irrad} = 0.25L_{\ast} ,
\end{equation}
where L$_{\ast}$ is the source luminosity.  
The maximum luminosity generated by viscous dissipation of
accretion within the disk is
\begin{equation}
L_{acc} = \frac{G M_{\ast} \dot{M}}{2 R_{\ast}},
\end{equation}
where M$_{\ast}$ and R$_{\ast}$ are the mass and radius of the central object, 
and $\dot{M}$ is the accretion rate for material 
moving onto the central object from the disk.

Modeling H$\alpha$ emission, \citet{muzerolle05} find low accretion rates 
for brown dwarfs in Taurus and Chamaeleon I with masses of 25 M$_{\rm J}$ 
($\dot{M}$ $\sim$ 5 $\times$ 10$^{-12}$ M$_{\odot}$ yr$^{-1}$). For 
objects of 25 M$_{\rm J}$ on the 1 Myr isochrone, 
\citet{baraffe03} predict R$_{\ast}$ $\simeq$ 
0.40 R$_{\odot}$, log(L$_{\ast}$/L$_{\odot}$) $\simeq$ -2.2.  
Using these values along with the equations above, we calculate 
L$_{irrad}$/L$_{acc}$ $\sim$320.  The mass accretion rate depends strongly 
on the stellar mass, with $\dot{M}$ $\propto$ M$^{2.1}$ for observations 
down to 25 M$_{\rm J}$ \citep{muzerolle05}.
If this relationship holds to lower masses, we would expect a 10 M$_{\rm J}$ object 
to have $\dot{M}$ $\sim$ 7 $\times$ 10$^{-13}$ M$_{\odot}$ yr$^{-1}$.  
The \citet{baraffe03} models predict R$_{\ast}$~$\simeq$~0.30~R$_{\odot}$
and log(L$_{\ast}$/L$_{\odot}$)~$\simeq$~-2.7 for a 1 Myr old, 10 M$_{\rm J}$ object
leading to a ratio of radiative luminosity to luminosity from viscous 
dissipation of accretion, 
L$_{irrad}$/L$_{acc}$ $\approx$1200.
Due to the smaller gravitational force experienced by their disks, 
brown dwarf disks should 
be highly flared, with scale heights up to 3 times those found for 
classical T Tauri stars \citep{walker04}.  In addition, if the disks 
have inner holes, then the inner radius of the disk 
would be the relevant radius in 
equation 2.  Best-fit models of the excess emission around 
two low mass brown dwarfs, OTS44 \citep{luhman05} and GY11 \citep{natta01}, 
indicate that low mass brown dwarfs can have inner radii of 3 R$_{\ast}$.   
Thus, our estimate of L$_{irrad}$/L$_{acc}$ is likely a lower limit.  
The observational evidence to date suggests that 
accretion should not play a major role in the heating of 
circumstellar material around low mass brown dwarfs.

\subsection{Excess vs. Stellar Luminosities}
\begin{figure}
\epsscale{0.8}
\plotone{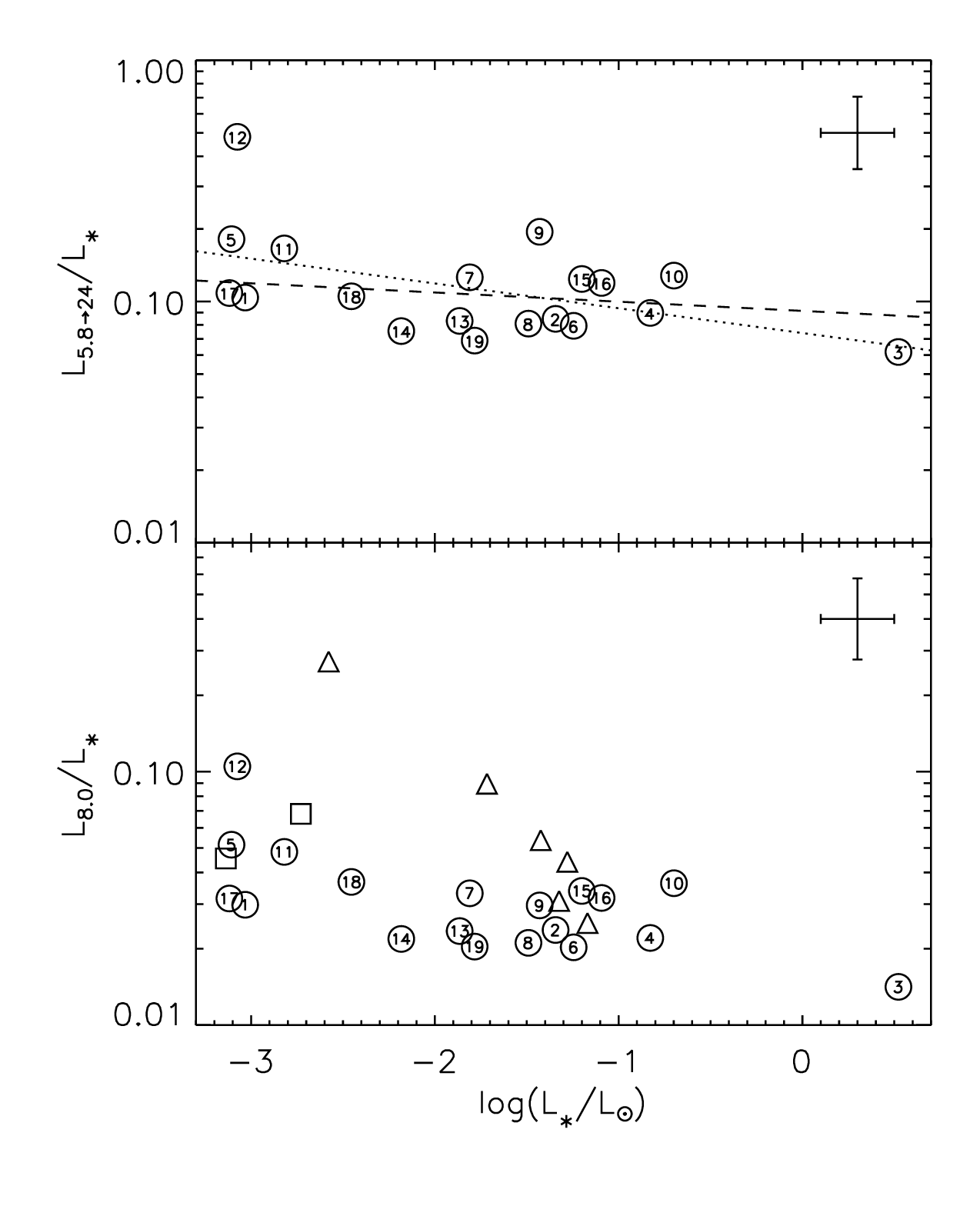}
\caption[1]
{\label{disk_vs_star} The ratio of excess emission to central source 
luminosity increases for lower luminosity sources.  
Upper Panel: Luminosity of the central source vs. the ratio of the IRAC3 
through MIPS1 excess luminosity to the central source luminosity.  
The dotted line shows a linear fit to our data points and the dashed 
line shows a linear fit excluding sources \#12 and \#3. Typical uncertainties 
(shown in the upper right) are 0.2 dex in luminosity and 0.15 dex in the excess to central source luminosity (see \S 3.2 for a discussion of uncertainties).
Lower Panel: Luminosity of the central source vs. the ratio of IRAC4 excess 
luminosity to the central source luminosity. The triangles show sources having
{\it i}-band magnitudes from 
\citet{natta02} and the squares show Cha~1109-7734 and OTS44 \citep{luhman05b,luhman05}.  
We calculated the luminosities and 
excesses using the method outlined in \S 3.2.  For the \citet{natta02} 
sources, we obtained near-IR fluxes from the 2MASS catalog and 
IRAC fluxes from the c2d survey of the Ophiuchus 
cloud core (Allen et al. 2006, in preparation).}
\end{figure}

The availability of fluxes over a broad range of near and mid-IR wavelengths 
for all the objects in our sample permits us to compare stellar and excess 
luminosities without the usual uncertainties introduced by using bolometric 
corrections to estimate the stellar luminosity or by 
measuring the excess at wavelengths where the object's photosphere dominates 
the emission.  
The upper panel of 
Figure \ref{disk_vs_star} shows the ratio of the 5.8 through 24 $\mu$m 
excess luminosity to the stellar (or central object) 
luminosity versus stellar luminosity, along with linear fits to 
all of our data points (dotted line) and excluding sources \#12 and \#3 
(dashed line).  
The ratio of excess to stellar luminosities is remarkably 
constant across three orders of magnitude in stellar luminosity. 
A linear fit (ordinary least-squares regression) 
to the data excluding sources \#12 and \#3 (two obvious outliers) 
shows that the ratio of excess to stellar 
luminosity scales as L$^{-a}$, where $a=0.04 \pm 0.04$.  
The mean 5.8--24~$\mu$m excess to stellar luminosity ratio for our sample 
(excluding sources \#12 and \#3) is 0.12 $\pm$ 30\%.  
The mean 5.8--24~$\mu$m excess to stellar luminosity ratio is lower than the 
minimum reprocessing luminosity (Equation 1), because we are not including 
flux longward of 24 $\mu$m, where much of the energy from the disk is emitted.
In the lower panel of Figure \ref{disk_vs_star}, we try to 
determine if the roughly 
constant ratio of excess to stellar luminosity holds for 
other samples of young, low-mass objects.  
Since a consistent data set of the disk luminosity is not available for 
other very low luminosity brown dwarfs, we plot the ratio of excess 
8.0~$\mu$m luminosity to the photospheric luminosity 
and include results for the \citet{natta02} and 
\citet{luhman05b,luhman05} brown dwarfs with disks.
Most of these additional sources 
are consistent with the constant ratio of excess to 
stellar luminosities that we find for higher luminosity sources.  The 
results for OTS44 \citep{luhman05} and the 2 lowest luminosity sources 
from \citet{natta02} show enhanced excess flux at 8~$\mu$m, consistent with 
a possible trend toward higher ratios of excess to stellar luminosity for 
very low luminosity brown dwarfs.
The constant ratio of excess to stellar luminosity for our 
sources is remarkable.  
{\it It implies that, whatever the details of the disk structure may be, the 
inner disks of young stars, brown dwarfs and sub-brown dwarfs intercept 
and reprocess about the same fraction of radiation from the central object. }

\section{Modeling the SEDs}
\begin{figure}
\hspace{1cm}
\includegraphics[angle=90,width=6in]{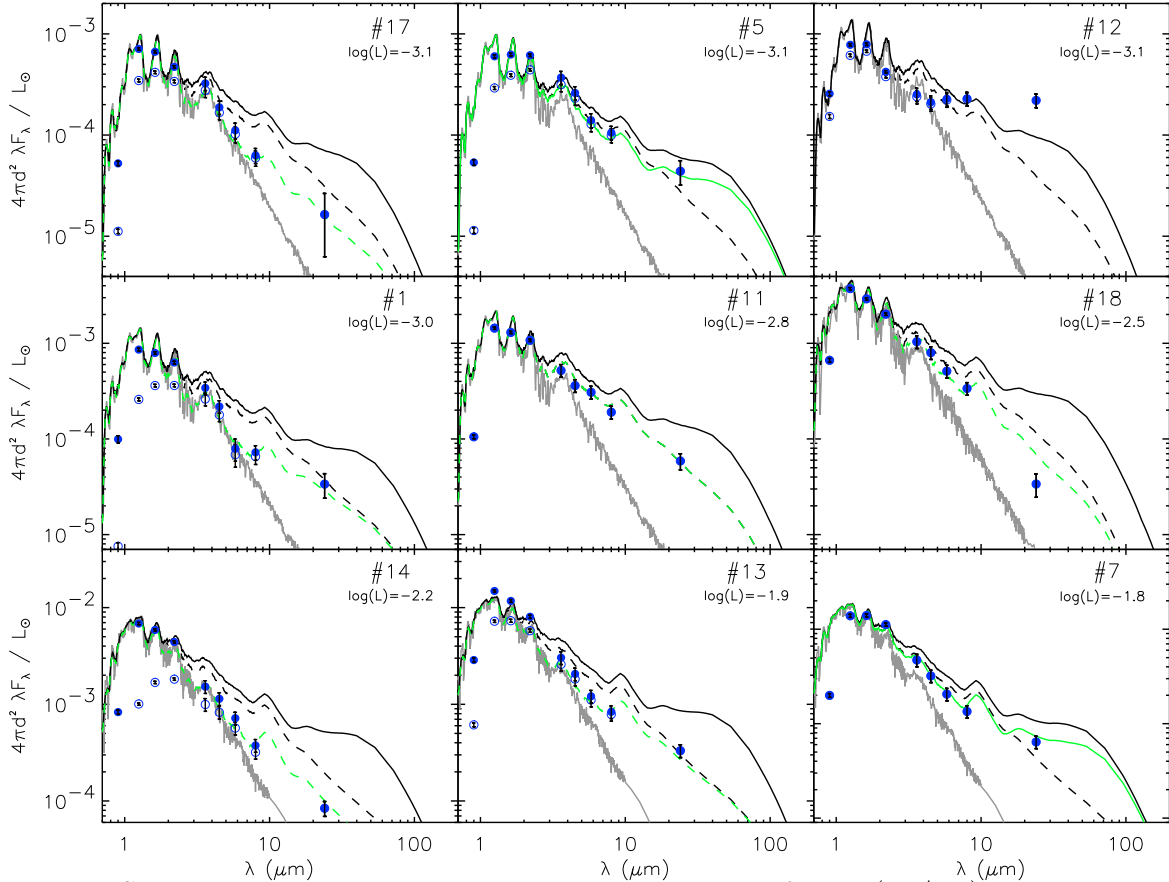}
\figcaption{SEDs with stellar atmosphere and disk models for $\rm{log} (L_{\ast}/L_{\odot}){\le}-1.8$. In each panel, the blue open and filled circles show the observed and dereddened data, respectively.  The grey solid line shows the SED of the stellar atmosphere model \citep{allard01,pheonix}.  The stellar atmospheres shown here have the values of T$_{eff}$ and log($g$) listed in Table \ref{modparams}, and are superposed (not fit) onto the data for the cloud distances listed in Table \ref{clouds}.  The black solid line is the predicted SED of a face-on, flared disk with an inner radius equal to the stellar radius ($R_i=R_{\ast}$).  The black dashed line is the predicted SED of a face-on, flat disk with $R_i=R_{\ast}$. The green line denotes the best fit to the data and is solid for flared disk models and dashed for flat disk models, as listed in Table~\ref{modparams}.   Source $\#$11 is best-fit by a flat, face-on disk with $R_i=R_{\ast}$.  Increased inner radii are required for $\#$1 ($R_i=4$~$R_{\ast}$) and $\#$13 ($R_i=3$~$R_{\ast}$).  Inclined disks with $R_i=R_{\ast}$ are required for $\#$5, $\#$7, and $\#$18 ($i=60^o$). Inclined disk ($i=60^o$) with large inner radii ($R_i=3$~$R_{\ast}$) are required for $\#$14 and $\#$17.
%\vspace{0.5cm}
\label{fig:sed1} }
\end{figure}

\begin{figure}
\hspace{1cm}
\includegraphics[angle=90,width=6in]{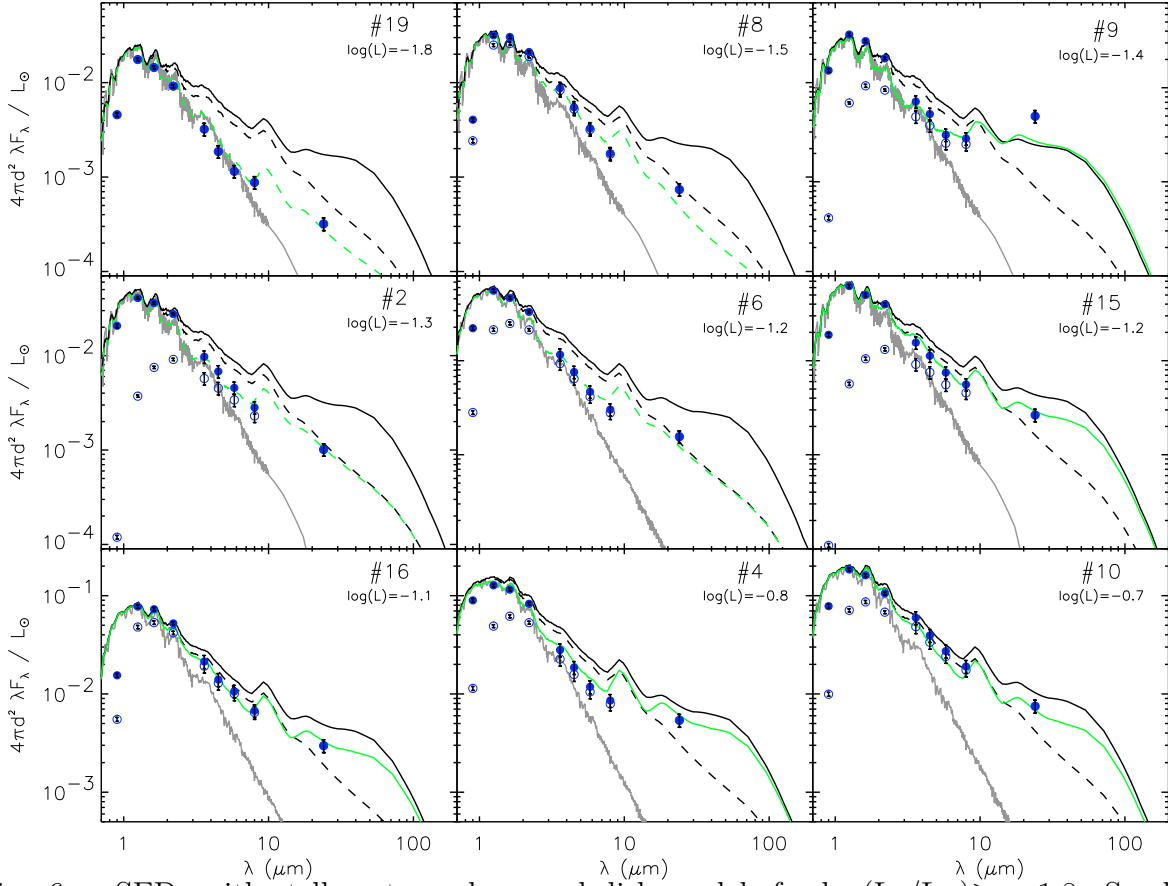}
\figcaption{SEDs with stellar atmosphere and disk models for $\rm{log} (L_{\ast}/L_{\odot}){\ge}-1.8$. Symbols and lines are as in Figure~\ref{fig:sed1}.   Increased inner radii ($R_i=3$~$R_{\ast}$) are required for $\#$2 and $\#$6.  Inclined disks ($i=60^o$) with $R_i=R_{\ast}$ are required for $\#$8, $\#$10, $\#$15, and $\#$16. Inclined disks ($i=60^o$) with large inner radii are required for $\#$4 ($R_i=3$~$R_{\ast}$), $\#$9 ($R_i=10$~$R_{\ast}$) and $\#$19 ($R_i=5$~$R_{\ast}$).
%\vspace{0.5cm}
\label{fig:sed2} }
\end{figure}

In Figures~\ref{fig:sed1}--\ref{fig:sed2}, we show the SEDs of 18 of the 19 
sources in our sample with detections in all bands  (I to MIPS1).  Source 
$\#$3 is not modeled as the luminosity of this source lies outside of the 
range covered by the \citet{baraffe98} models.
 If we match the source luminosities to the 1 Myr 
(for Ophiuchus and Lupus~I) 
and 3 Myr (for Chameleon II) isochrones of \citet{baraffe03} or 
\citet{baraffe98}, then Figure~\ref{fig:sed1} includes sources with 
inferred masses of 6--50 $M_J$ ($\rm{log} (L_{\ast}/L_{\odot}){\le}-1.8$).
and Figure~\ref{fig:sed2} includes sources with masses 
of 50--350 $M_J$ ($\rm{log} (L_{\ast}/L_{\odot}){\ge}-0.7$). 

Table~\ref{modparams} lists the luminosity, effective temperature, 
mass, radius and gravity of the evolutionary model 
\citep{baraffe03, baraffe98} with L$_{model}$ closest to 
L$_{source}$ at the age of the cloud (Table \ref{clouds}).  The stellar 
atmosphere models \citep[][grey lines in 
Figures~\ref{fig:sed1} and \ref{fig:sed2}]{allard01,pheonix} are for the 
T$_{eff}$ listed in Table \ref{modparams} rounded to the nearest 50~K, and a 
gravity, log($g$)~=~3.5, appropriate for 1--3 Myr objects.
For objects near or above the stellar limit (i.e., for $T_{eff}\ge3000$ K), we use the PHOENIX model atmospheres for low mass stars \citep{pheonix}\footnote{PHOENIX models were obtained from {\url http://www.hs.uni-hamburg.de/EN/For/ThA/phoenix}}.   For substellar objects ($T_{eff}\le$~3000~K), we use AMES-dusty models, which 
include dust formation in the chemical equilibrium calculations and the 
contribution of dust opacities to the total optical depth.  The 
atmospheric models shown in Figure~\ref{fig:sed1}--\ref{fig:sed2} have 
not been scaled to fit the data, but are simply multiplied by a dilution 
factor $f=(R_{\ast}/d)^2$, using the distances from Table \ref{clouds}.
The coincidence between the atmospheric models 
and the flux in the 0.9--3.6 $\mu$m range is generally quite good, with 
clear excesses visible beyond 5.0 $\mu$m.  The model atmospheres generally 
predict bluer near-IR colors than are observed.

In order to interpret the observed excesses, we compare the observed SEDs 
to model predictions for passive, irradiated disks 
\citep[CGPLUS;][]{dullemond01}.  CGPLUS is based on the 
\citet[][hereafter CG97]{chiang97} model of 
a flared two-layer disk. The CG97 model assumes vertical hydrostatic 
equilibrium, resulting in a flared structure. In this model, the dust and gas 
are well mixed and no dust settling has occurred. The optically 
thin disk surface layer is irradiated directly by the central object.  
Superheated dust grains in the surface layer re-radiate into the disk 
interior and thus regulate the interior disk temperature.  CGPLUS is 
a modified version of CG97, including the addition of a puffed up 
inner disk wall, the height of which is calculated self-consistently.  
Although CGPLUS has not previously been applied to brown dwarfs, similar 
models have proven to be adequate representations 
of disks around sub-stellar objects \citep{natta01}.

For simplicity, we begin with disk model parameters very similar to those of 
T~Tauri stars.  For a source of estimated mass $M_{\ast}$, we adopt a disk 
mass of 0.03 ${M_{\ast}}$ and a fixed outer disk radius of 5 AU. Increasing
the disk outer radius does not change the SED shortward of ~25 $\mu$m, and
thus our observations cannot constrain this outer radius.  However, for our 
lowest mass objects, the disk scale heights become unreasonably large at 
R $\gtrsim 5$~AU, such that the material is no longer gravitationally bound 
to the central source. The 
surface density of the disk varies as $R^{-1}$ and opacities are from 
\citet{laor93}. All sources are modeled 
with flat and flared face-on disks ($i=0^o$) with a disk inner radius 
equal to the stellar radius ($R_i={R_{\ast}}$).  Where necessary, 
additional models with larger inner disk radii and inclinations are 
calculated. 

We test the effects of the disk inclination, disk flaring and the size 
of the inner disk hole on the strength/shape of the excess.  
In general, SEDs of flared inclined disks (e.g. $i=60^o$) are quite similar 
to SEDs of flat face-on disks for $\lambda \le 24$~$\mu$m, and our 
IRAC observations often cannot distinguish between these two models.  
For flat inclined disks (e.g. $i=60^o$), the flux falls off more rapidly as a 
function of wavelength. 
If the inner radius is increased to $R_i>R_{\ast}$, then the flux in 
the IRAC bands 
becomes more photospheric.  For flat disks with $R_i>R_{\ast}$, 
the flux beyond $\lambda\approx15$~$\mu$m is similar to a flat disk with 
$R_i=R_{\ast}$. For flared disks with large inner holes ($R_i>R_{\ast}$), 
the flux beyond 
$\lambda\approx15$~$\mu$m increases because the surface layer of the disk is 
warmer at larger radii.  

The best fit models are overlaid in green in Figures~\ref{fig:sed1} and 
\ref{fig:sed2} and the parameters for these models are 
listed in Table~\ref{modparams}.
The SEDs of all but one of the sources in this sample
have power-law slopes of the SED, $a$, in the range of 1.2--2 
(where ${\nu}F_{\nu}\propto{\nu}^{a}$) and can be fit well by models 
of flared and/or flat disks (see Figure~\ref{fig:sed2}).  The exception is $\#$12, for which the photospheric model fits well, but the excess at $\lambda\ge5.8$~$\mu$m 
is too large to be fit with the models presented here.  An actively accreting 
disk or non-disk geometry (such as a tenuous dust envelope), might be 
able to explain the mid-IR excess for source \#12, but a quantitative analysis 
of these scenarios is beyond the scope of this paper.
Both flared disks with large 
inclinations $i\ge60^o$ and face-on flat disks fit sources 
with $a\approx{4/3}$ (\#5, \#7, \#11 and \#16).  
The distinction between flared, inclined disks and flat, face-on disks relies 
on the accuracy of one data point, our 24$\mu$m flux measurement, and is 
thus the least constrained parameter of our model fits.
About half of the 
 SEDs show extremely low excesses in the 3.6--5.8~$\mu$m 
range and the SEDs can be fit best with a disk model possessing a larger 
inner radius ($R_i=$2--10 ${R_{\ast}}$; e.g., $\#$1 and $\#$3).

The sub-brown dwarfs in our sample to which we can fit disk models 
(\#1, \#5, \#11, \& \#17) are fit by 3 flat disks and 1 flared disk, 
and a range of inner disk radii 
($R_i=1-4 R_{\ast}$).  The average inner disk radius we fit to sub-brown dwarfs is 
$R_i\simeq 2 R_{\ast}$.  
The brown dwarfs in our sample (\#7, \#8, \#9, \#13, \#14, \#18, and \#19) 
have a similar distribution 
of flared vs. flat disks (2 out of 7 are flared) and disk inner radii (with the exception of \#9, $R_i \simeq 2 R_{\ast}$ on average) as 
the sub-brown dwarfs. 
The young stars in our sample have an average inner disk 
radii, $R_i \simeq 2 R_{\ast}$, but have a larger fraction of flared disks 
(4 out of 6 are flared) than found 
for the brown dwarfs in our sample.  
The similarity of the disk properties for young stars, 
brown dwarfs and sub-brown dwarfs, along with the constant ratio of excess to 
stellar luminosities (\S4.3), indicates that disks across the mass range 
probably form and evolve in similar ways. 

\section{Conclusions}

In this paper we have presented a sample of young objects with luminosities 
ranging from 0.5 $>$ log(L$_{\ast}$/L$_{\odot}$) $>$ -3.1 in the 
Chamaeleon~II, Ophiuchus, and Lupus~I star-forming clouds.  
The lowest luminosity 
sources in our sample (log(L$_{\ast}$/L$_{\odot}$) $<$ -3.0) have inferred 
masses of 6--12~M$_{\rm J}$ based on the 1 and 3 Myr isochrones of 
\citet{baraffe03}.
The 5.8--24 $\mu$m fluxes of these objects show evidence of excess emission 
above the photosphere, presumably from a circumstellar disk.  The ratio of 
5.8--24 $\mu$m excess to stellar luminosity is constant 
(to within 30\%) 
over three decades in luminosity.

Most of the near-IR fluxes of our objects agree with the 
predictions of stellar atmosphere models \citep{allard01}.  
Given that our choices of 
T$_{eff}$ and log(g) are based on the measured source luminosities and 
model isochrone predictions of stellar 
luminosities at the ages of Chamaeleon~II (3 Myr), Lupus~I and Ophiuchus 
(1 Myr), rather than to fits of the spectral energy distributions, 
the agreement is remarkable.  
In general, the stellar atmosphere models tend to overestimate the I-band 
flux, and predict bluer near-IR colors.
The origin of the discrepancy between 
the observations and model spectra can be 
explored once spectra are obtained for these objects.

Theoretical models of passive irradiated disks fit the excess emission 
of all but one of the objects in our sample.  We find that flared, inclined 
disks have similar SEDs to flat, face-on disks in the IRAC wavelengths, but our
observations at 24~$\mu$m can distinguish between the two in most cases.  
We fit a similar distribution 
of flared vs. flat disks, disk inclination angles, and disk inner radii 
to the young stars, brown dwarfs, and sub-brown dwarfs in our sample.

\acknowledgments

We would like to thank Cornelis Dullemond for providing his 
models in electronic form and working with us to adapt/test his CGPLUS models
for disks around very low mass objects.  
We are grateful to  Juan Alcal{\' a}, Lori Allen, Isabelle Baraffe, 
Neal Evans, and Dave Koerner for comments on the 
manuscript.  Support for JEK-S was provided by the Spitzer 
Space Telescope Postdoctoral Fellowship Program, under award 1256316.
Support for this work, part of the Spitzer Legacy Science
Program, was provided by NASA through contracts 1224608 and 1256316
issued by the Jet Propulsion Laboratory, California Institute of
Technology, under NASA contract 1407.

\clearpage

%%%%%%%%%%%%%%%%% Table 1 %%%%%%%%%%%%%%%%%%%%%%%%%%%%%%%
\begin{deluxetable}{lcrrrrrrrrr}
\tablecolumns{10}
\footnotesize
\tablecaption{Observations}
\tablewidth{0pt}
\tablehead{
\colhead{Cloud}                  &
\colhead{Area\tablenotemark{a}}             &
\multicolumn{9}{c}{10$\sigma$ Limit\tablenotemark{b}} \\
\colhead{} &
\colhead{} &
\colhead{I} &
\colhead{J} &
\colhead{H} &
\colhead{Ks} &
\colhead{[3.6]} &
\colhead{[4.5]} &
\colhead{[5.8]} &
\colhead{[8.0]} &
\colhead{[24.0]}\\
\colhead{} &
\colhead{sq. arcmin} &
\multicolumn{8}{c}{mag} &
\colhead{mJy} \\
}
\startdata
Chamaeleon~II & 2200 & 22.9 & 19.7 & 19.1 & 18.6 & 17.5 & 16.8 & 14.7 & 14.4 & 0.9\\
Ophiuchus     & 1700 & 23.5 & 20.2 & 19.6 & 18.9 & 17.1 & 16.4 & 14.1 & 13.7 & 0.7\\
Lupus~I       & 2100 & 23.3 & 20.0 & 19.4 & 18.8 & 17.4 & 16.5 & 14.3 & 13.9 & 0.9
\enddata
\tablenotetext{a}{The area covered in all 4 near-IR bands}
\tablenotetext{b}{Calculated from the average flux of extracted sources in each band with signal-to-noise between 9.8 and 10.3}
\label{limits}
\end{deluxetable}

%%%%%%%%%%%%%%%%% Table 5 %%%%%%%%%%%%%%%%%%%%%%%%%%%%%%%
\begin{deluxetable}{lccc}
\tablecolumns{6}
\footnotesize
\tablecaption{Adopted Cloud Parameters}
\tablewidth{0pt}
\tablehead{
\colhead{Cloud}                  &
\colhead{distance\tablenotemark{a}}             &
\colhead{modulus} &
\colhead{Age\tablenotemark{b}}   \\
\colhead{} &
\colhead{pc} &
\colhead{mag} &
\colhead{Myr} \\
}
\startdata
Chamaeleon~II & 178$\pm$18\tablenotemark{c} & 6.25 & 3 \\
Ophiuchus     & 125$\pm$25\tablenotemark{d} & 5.48 & 1 \\
Lupus~I       & 150$\pm$20\tablenotemark{e} & 5.88 & 1 
\enddata
\tablenotetext{a}{distances as adopted by the c2d team (Neal Evans, private communication)}
\tablenotetext{b}{See \S 3.3 and references therein}
\tablenotetext{c}{\citet{whittet97}}
\tablenotetext{d}{\citet{degeus89}}
\tablenotetext{e}{Comeron, F., in preparation}
\label{clouds}
\end{deluxetable}

%%%%%%%%%%%%%%%%% Table 2 %%%%%%%%%%%%%%%%%%%%%%%%%%%%%%%
%\begin{rotate}{90}
\begin{deluxetable}{r|r|r|rrrrrrrr}
\rotate
\tablecolumns{11}
\tabletypesize{\scriptsize}
\tablecaption{Photometry of our Sources}
\tablewidth{0pt}
\tablehead{
\colhead{\#}     &
\multicolumn{1}{c}{$\alpha$(J2000)}                  &
\multicolumn{1}{c}{$\delta$(J2000)}             &
\colhead{I}            &
\colhead{J}            &
\colhead{H}            &
\colhead{Ks}            &
\colhead{[3.6]}            &
\colhead{[4.5]}            &
\colhead{[5.8]}            &
\colhead{[8.0]}            \\
}
\startdata
 1 & 12 57 58.7& -77 01 19.5& $22.61\pm0.10$ & $17.88\pm0.04$ & $16.80\pm0.03$ & $15.98\pm0.03$ & $14.85\pm0.16$ & $14.55\pm0.16$ & $14.84\pm0.28$ & $13.87\pm0.19$\\
 2\tablenotemark{a} & 12 58 06.7& -77 09 09.5& $19.61\pm0.05$ & $14.99\pm0.03$ & $13.50\pm0.03$ & $12.48\pm0.03$ & $11.49\pm0.16$ & $11.04\pm0.16$ & $10.60\pm0.16$ & $10.01\pm0.16$\\
 3\tablenotemark{b} & 13 00 59.3& -77 14 02.7& $16.26\pm0.05$ & $11.48\pm0.04$ & $ 9.55\pm0.04$ & $ 7.92\pm0.04$ & $ 6.84\pm0.16$ & $ 6.48\pm0.17$ & $ 6.20\pm0.16$ & $ 5.97\pm0.16$\\
 4\tablenotemark{c} & 13 04 24.9& -77 52 30.3& $14.66\pm0.05$ & $12.19\pm0.04$ & $11.22\pm0.04$ & $10.57\pm0.04$ & $10.00\pm0.16$ & $ 9.68\pm0.16$ & $ 9.38\pm0.16$ & $ 8.67\pm0.16$\\
 5 & 13 05 40.8& -77 39 58.2& $22.16\pm0.07$ & $17.75\pm0.04$ & $16.71\pm0.03$ & $15.77\pm0.03$ & $14.64\pm0.16$ & $14.26\pm0.16$ & $14.16\pm0.17$ & $13.42\pm0.17$\\
 6\tablenotemark{d} & 13 07 18.1& -77 40 52.9& $16.18\pm0.05$ & $13.09\pm0.04$ & $12.20\pm0.04$ & $11.56\pm0.04$ & $10.96\pm0.16$ & $10.65\pm0.16$ & $10.38\pm0.16$ & $ 9.80\pm0.16$\\
 7\tablenotemark{e} & 13 08 27.1& -77 43 23.3& $16.55\pm0.05$ & $13.56\pm0.04$ & $12.83\pm0.04$ & $12.26\pm0.04$ & $11.71\pm0.16$ & $11.41\pm0.16$ & $11.15\pm0.16$ & $10.59\pm0.16$\\
 8 & 16 21 42.0& -23 13 43.2& $15.58\pm0.05$ & $12.15\pm0.04$ & $11.39\pm0.04$ & $10.92\pm0.04$ & $10.33\pm0.16$ & $10.10\pm0.16$ & $ 9.91\pm0.16$ & $ 9.54\pm0.16$\\
 9 & 16 21 48.5& -23 40 27.3& $17.50\pm0.05$ & $13.55\pm0.03$ & $12.37\pm0.04$ & $11.68\pm0.03$ & $10.90\pm0.16$ & $10.41\pm0.16$ & $10.14\pm0.16$ & $ 9.15\pm0.16$\\
10\tablenotemark{f} & 16 22 25.0& -23 29 55.4& $14.04\pm0.05$ & $11.02\pm0.04$ & $10.08\pm0.04$ & $ 9.53\pm0.04$ & $ 8.41\pm0.16$ & $ 8.09\pm0.16$ & $ 7.71\pm0.16$ & $ 7.04\pm0.16$\\
11 & 16 22 25.2& -24 05 15.6& $18.98\pm0.05$ & $15.24\pm0.03$ & $14.64\pm0.03$ & $14.03\pm0.03$ & $13.33\pm0.16$ & $13.01\pm0.16$ & $12.44\pm0.17$ & $11.94\pm0.17$\\
12 & 16 22 30.2& -23 22 24.0& $18.58\pm0.05$ & $16.17\pm0.03$ & $15.35\pm0.03$ & $15.17\pm0.03$ & $14.17\pm0.16$ & $13.64\pm0.17$ & $12.80\pm0.17$ & $11.76\pm0.16$\\
13 & 16 22 44.9& -23 17 13.4& $17.07\pm0.05$ & $13.50\pm0.03$ & $12.76\pm0.04$ & $12.21\pm0.04$ & $11.59\pm0.16$ & $11.25\pm0.16$ & $11.05\pm0.16$ & $10.41\pm0.16$\\
14 & 16 23 05.8& -23 38 17.8& $21.22\pm0.05$ & $15.64\pm0.03$ & $14.36\pm0.03$ & $13.46\pm0.03$ & $12.63\pm0.16$ & $12.12\pm0.16$ & $11.78\pm0.16$ & $11.39\pm0.16$\\
15 & 16 23 15.7& -23 43 00.4& $18.94\pm0.05$ & $13.76\pm0.04$ & $12.37\pm0.04$ & $11.31\pm0.04$ & $10.22\pm0.16$ & $ 9.72\pm0.16$ & $ 9.30\pm0.16$ & $ 8.50\pm0.16$\\
16 & 16 23 36.1& -24 02 20.9& $14.68\pm0.05$ & $11.44\pm0.04$ & $10.61\pm0.05$ & $10.06\pm0.04$ & $ 9.41\pm0.17$ & $ 9.13\pm0.17$ & $ 8.66\pm0.16$ & $ 8.12\pm0.16$\\
17 & 15 39 27.3& -34 48 44.0& $21.81\pm0.05$ & $17.19\pm0.03$ & $16.27\pm0.03$ & $15.69\pm0.03$ & $14.41\pm0.16$ & $14.25\pm0.16$ & $14.04\pm0.19$ & $13.61\pm0.19$\\
18 & 15 41 40.8& -33 45 18.8& $17.38\pm0.05$ & $14.61\pm0.03$ & $14.16\pm0.03$ & $13.75\pm0.03$ & $12.97\pm0.16$ & $12.55\pm0.16$ & $12.29\pm0.17$ & $11.72\pm0.16$\\
19 & 15 44 57.9& -34 23 39.3& $15.28\pm0.05$ & $12.93\pm0.04$ & $12.42\pm0.04$ & $12.10\pm0.04$ & $11.75\pm0.16$ & $11.63\pm0.16$ & $11.41\pm0.16$ & $10.69\pm0.16$
\enddata
\tablenotetext{a}{Previously identified as ISO-ChaII-13 \citep{persi03}}
\tablenotetext{b}{Previously identified as ISO-ChaII-54 \citep{persi03}, 
C48-DENIS \citep{vuong01}, and CHIIXR10 \citep{alcala00}}
\tablenotetext{c}{Previously identified as Sz52 \citep{schwartz77}}
\tablenotetext{d}{Previously identified as C62-DENIS \citep{vuong01}}
\tablenotetext{e}{Previously identified as C66-DENIS \citep{vuong01}}
\tablenotetext{f}{Previously identified as WSB 14 \citep{wilking87}}
\label{sources}
\end{deluxetable}
%\end{rotate}

%%%%%%%%%%%%%%%%% Table 3 %%%%%%%%%%%%%%%%%%%%%%%%%%%%%%%
\begin{deluxetable}{rrrrrr}
\tablecolumns{6}
\tabletypesize{\footnotesize}
\tablecaption{Mid-IR Fluxes}
\tablewidth{0pt}
\tablehead{
\colhead{\#}  &
\colhead{$F_{3.6}$}            &
\colhead{$F_{4.5}$}            &
\colhead{$F_{5.8}$}            &
\colhead{$F_{8.0}$}            &
\colhead{$F_{24}$}            \\
\colhead{}            &
\multicolumn{5}{c}{mJy} \\
}
\startdata
 1 & $  0.32\pm  0.05$ & $  0.27\pm 0.04$ & $  0.14\pm 0.04$ & $  0.18\pm 0.03$ & $  0.28\pm 0.08$\\
 2 & $  7.00\pm  1.05$ & $  6.91\pm 1.04$ & $  6.69\pm 1.00$ & $  6.25\pm 0.94$ & $  8.26\pm 1.24$\\
 3 & $508.11\pm 76.28$ & $460.99\pm70.06$ & $385.71\pm57.91$ & $258.41\pm39.03$ & $492.73\pm74.01$\\
 4 & $ 27.66\pm  4.15$ & $ 24.13\pm 3.62$ & $ 20.71\pm 3.11$ & $ 21.50\pm 3.23$ & $ 43.91\pm 6.59$\\
 5 & $  0.39\pm  0.06$ & $  0.35\pm 0.05$ & $  0.25\pm 0.04$ & $  0.27\pm 0.04$ & $  0.36\pm 0.10$\\
 6 & $ 11.44\pm  1.72$ & $  9.86\pm 1.48$ & $  8.20\pm 1.23$ & $  7.57\pm 1.14$ & $ 12.73\pm 1.91$\\
 7 & $  5.77\pm  0.87$ & $  4.90\pm 0.74$ & $  4.05\pm 0.61$ & $  3.65\pm 0.55$ & $  5.22\pm 0.79$\\
 8 & $ 20.54\pm  3.08$ & $ 16.32\pm 2.45$ & $ 12.68\pm 1.90$ & $  9.65\pm 1.45$ & $ 12.18\pm 1.83$\\
 9 & $ 12.16\pm  1.83$ & $ 12.29\pm 1.85$ & $ 10.27\pm 1.54$ & $ 13.81\pm 2.07$ & $ 82.18\pm12.34$\\
10 & $119.82\pm 18.10$ & $104.04\pm15.62$ & $ 96.43\pm14.48$ & $ 96.46\pm14.48$ & $123.64\pm18.57$\\
11 & $  1.29\pm  0.19$ & $  1.12\pm 0.17$ & $  1.23\pm 0.19$ & $  1.05\pm 0.16$ & $  0.97\pm 0.19$\\
12 & $  0.59\pm  0.09$ & $  0.63\pm 0.10$ & $  0.89\pm 0.14$ & $  1.25\pm 0.19$ & $  3.64\pm 0.56$\\
13 & $  6.41\pm  0.96$ & $  5.66\pm 0.85$ & $  4.43\pm 0.67$ & $  4.34\pm 0.65$ & $  5.48\pm 0.83$\\
14 & $  2.47\pm  0.37$ & $  2.56\pm 0.38$ & $  2.27\pm 0.34$ & $  1.76\pm 0.27$ & $  1.39\pm 0.24$\\
15 & $ 22.70\pm  3.41$ & $ 23.32\pm 3.50$ & $ 22.23\pm 3.34$ & $ 25.13\pm 3.77$ & $ 43.73\pm 6.57$\\
16 & $ 47.66\pm  7.24$ & $ 40.00\pm 6.08$ & $ 40.09\pm 6.02$ & $ 35.66\pm 5.35$ & $ 49.18\pm 7.38$\\
17 & $  0.48\pm  0.07$ & $  0.36\pm 0.05$ & $  0.28\pm 0.05$ & $  0.23\pm 0.04$ & $  0.19\pm 0.12$\\
18 & $  1.79\pm  0.27$ & $  1.72\pm 0.26$ & $  1.42\pm 0.22$ & $  1.29\pm 0.19$ & $  0.39\pm 0.11$\\
19 & $  5.54\pm  0.83$ & $  4.01\pm 0.60$ & $  3.19\pm 0.48$ & $  3.35\pm 0.50$ & $  3.67\pm 0.56$
\enddata
\label{iracsources}
\end{deluxetable}

%%%%%%%%%%%%%%%%% Table 4 %%%%%%%%%%%%%%%%%%%%%%%%%%%%%%%
\begin{deluxetable}{rrr}
\tablecolumns{3}
\tabletypesize{\footnotesize}
\tablecaption{Source Luminosities}
\tablewidth{0pt}
\tablehead{
\colhead{\#}             &
\colhead{A$_V$}            &
\colhead{log(L$_{\ast}$/L$_{\odot}$)}\\
}
\startdata
 1 &  5 & -3.0\\
 2 & 10 & -1.3\\
 3 & 13 &  0.5\\
 4 &  4 & -0.8\\
 5 &  3 & -3.1\\
 6 &  4 & -1.2\\
 7 &  0 & -1.8\\
 8 &  1 & -1.5\\
 9 &  7 & -1.4\\
10 &  4 & -0.7\\
11 &  0 & -2.8\\
12 &  1 & -3.1\\
13 &  3 & -1.9\\
14 &  8 & -2.2\\
15 & 10 & -1.2\\
16 &  2 & -1.1\\
17 &  3 & -3.1\\
18 &  0 & -2.5\\
19 &  0 & -1.8
\enddata
\label{luminosities}
\end{deluxetable}

%%%%%%%%%%%%%%%%% Table 6 %%%%%%%%%%%%%%%%%%%%%%%%%%%%%%%
\begin{deluxetable}{lccccc|cccc}
\tablewidth{0pt}
\tabletypesize{\footnotesize}
\tablecaption{Model parameters}
\tablehead{\colhead{ } & \colhead{L$_{model}$} & \colhead{T$_{eff}$} & \colhead{Mass} & \colhead{R$_{\ast}$} & \colhead{} & \colhead{R$_{i}$} & \colhead{$i$} & \colhead{ }   \\ 
\colhead{$\#$}      & \colhead{log(L$_{\ast}$/L$_{\odot}$)}  & \colhead{(K)} & \colhead{(M$_{\rm J}$)} & \colhead{(R$_{\odot}$)} & \colhead{Log $(g)$} & \colhead{(R$_{\ast}$)} & \colhead{($^o$)} & \colhead{geometry} }
\startdata
%# L_mod    Teff       M         R        g    R_i    i   geometry
1  & -2.99  & 2207  & 12    & 0.22   & 3.83 & 4   & 0  & flat   \\ 
2  & -1.31  & 2925  & 100   & 0.87   & 3.56 & 3   & 0  & flat   \\ 
%3  &  0.43  & 5000  & 1600  & 2.1924   & ...  & ...  & ... & ... \\ 
4  & -0.79  & 3395  & 350   & 1.17   & 3.84 & 3   & 60 & flared \\ 
5  & -3.14  & 2100  & 10    & 0.20   & 3.81 & 1   & 60 & flared \\  
6  & -1.20  & 3140  & 175   & 0.85   & 3.82 & 3   & 0  & flat \\  
7  & -1.77  & 2793  & 50    & 0.56   & 3.63 & 1   & 60  & flared   \\      
8  & -1.48  & 2853  & 70    & 0.75   & 3.68 & 1   & 60 & flat   \\      
9  & -1.45  & 2858  & 72    & 0.77   & 3.52 & 10  & 60 & flared \\
10 & -0.70  & 3193  & 200   & 1.46   & 3.40 & 1   & 60 & flared \\
11 & -2.81  & 2207  & 9     & 0.27   & 3.53 & 1   & 0 & flat \\
12\tablenotemark{a} & -3.00  & 2098  & 7     & 0.24   & 3.51 & ... & ... & ...   \\     
13 & -1.94  & 2746  & 40    & 0.48   & 3.68 & 3   & 0 & flat \\
14 & -2.17  & 2598  & 30    & 0.41   & 3.69 & 3   & 60 & flat   \\      
15 & -1.19  & 2856  & 100   & 1.05   & 3.39 & 1   & 60 & flared \\  
16 & -1.08  & 3023  & 110   & 1.05   & 3.43 & 1   & 60  & flared   \\ 
17 & -3.13  & 2004  & 6     & 0.28   & 3.50 & 3   & 60 & flat   \\ 
18 & -2.46  & 2400  & 15    & 0.34   & 3.54 & 1   & 60  & flat \\ 
19 & -1.64  & 2768  & 50    & 0.66   & 3.49 & 5   & 60 & flat 
\enddata
\tablecomments{ ${\rm L}_{model}$, T$_{eff}$, M$_{\ast}$, R$_{\ast}$ and Log $(g)$ are the luminosity, effective temperature, mass, radius, and gravity of the 
evolutionary model \citep{baraffe03, baraffe98} with L$_{model}$ closest to L$_{source}$ at the age of the cloud (\S3.3). The last three columns list the disk inner radius (R$_{i}$), inclination ($i$) and geometry for the CGPLUS models that best fit the data (shown as green lines in Figures~\ref{fig:sed1} and \ref{fig:sed2}).}
\tablenotetext{a} {For this source, the flux at $\lambda\ge5.8$~$\mu$m is too large to be fit by the models presented here.}
%\tablerefs{(1) \citealp{Lev88}; (2) \citealp{CoKu79}; (3) \citealp{FEOM}} 
\label{modparams}
\end{deluxetable}

\clearpage

\end{document}